\documentclass{aa} 
\usepackage{graphicx}
\usepackage{amsmath}
\usepackage{amssymb}
\usepackage{natbib}
\usepackage{hyperref}
\usepackage{booktabs}
\usepackage{multirow}
\usepackage{float}

\hypersetup{
	colorlinks=true,
	linkcolor=black,
	citecolor=blue,
	urlcolor=blue
}

\newcommand{\ez}{\mathbf{e}_{\rm z}}

\usepackage{amsmath}
\usepackage{graphicx}
\usepackage{txfonts}
\usepackage{hyperref}
\usepackage{color}

\begin{document} 

\title{Polar cap plasma loading and the morphology of pulsar $\gamma$-ray light curves} 


\author{J. P\'etri\inst{1}
          }

\institute{Universit\'e de Strasbourg, CNRS, Observatoire astronomique de Strasbourg, UMR 7550, F-67000 Strasbourg, France.\\
\email{jerome.petri@astro.unistra.fr}   
}

\date{Received ; accepted }

 
\abstract 
{The discovery of more than 300 $\gamma$-ray pulsars by the Fermi Large Area Telescope (Fermi-LAT) has revealed a rich diversity of light-curve morphologies that remains challenging to reproduce within purely magnetospheric emission models, suggesting that particle acceleration and radiation may occur, at least in part, in the striped-wind current sheet.}
{We compute a comprehensive atlas of pulsar $\gamma$-ray light curves based on the split-monopole current sheet geometry, with the goal of quantifying the role of pair plasma loading in shaping the observed pulse morphology.}
{The current-sheet surface is described analytically and assumed to be the dominant site of high-energy photon emission. Spatially dependent emissivity prescriptions are introduced through a minimal set of parameters that explicitly break the north-south and azimuthal symmetries of the emitting plasma.}
{With only a few physically motivated parameters, the model is able to reproduce a broad range of the observed $\gamma$-ray light-curve morphologies, including asymmetric, multi-peaked, and highly structured profiles.}
{The results show that asymmetric emissivity, possibly related to pair-plasma loading, provides a natural and economical mechanism capable of reproducing a substantial fraction of the observed diversity of $\gamma$-ray pulse profiles. The current-sheet scenario therefore appears to provide a promising and physically motivated framework for high-energy pulsar emission.}

\keywords{Magnetic fields -- Pulsars: general -- Gamma rays: stars -- Magnetospheres -- Radiation mechanisms: non-thermal -- Methods: analytical}

\maketitle

%

\section{Introduction}

The advent of the Fermi Large Area Telescope (Fermi/LAT) has profoundly transformed our understanding of pulsars at high energies (GeV), leading to the detection of more than 300 $\gamma$-ray pulsars and revealing an exceptional diversity of light-curve morphologies \citep{smith_third_2023}. The observed profiles encompass a wide range of characteristics, including sharp and widely separated peaks, asymmetric components, and complex multi-peaked structures. While geometric models based on magnetospheric emission zones, such as polar cap, slot gap, and outer gap scenarios, have successfully reproduced some general trends, they still encounter significant limitations in accounting for the full diversity of observed light curves within a unified and physically consistent framework.

In parallel, global simulations of pulsar magnetospheres, including force-free, resistive, and particle-in-cell approaches \citep{spitkovsky_time-dependent_2006, kalapotharakos_extended_2012, cerutti_modelling_2016}, have demonstrated the fundamental role of the equatorial current sheet that forms beyond the light cylinder. In these models, the striped-wind region naturally hosts magnetic reconnection and efficient particle acceleration, making it a prime candidate for high-energy emission \citep{coroniti_magnetically_1990, michel_magnetic_1994} \citep{petri_unified_2011, uzdensky_physical_2014}. This has led to a paradigm shift in which a substantial fraction of the observed $\gamma$-ray radiation is produced outside the light cylinder, in the dissipative current sheet of the pulsar wind.

Analytical striped-wind models have proven particularly valuable for establishing a direct link between wind geometry and observed emission properties. In particular, previous studies have shown that such models can reproduce key features of pulsar light curves, including peak separations, phase lags, and caustic formation \cite{petri_unified_2011, bai_modeling_2010}. Extensions incorporating more realistic emission prescriptions have further highlighted the importance of relativistic beaming and the current-sheet structure depending on the plasma resistivity \citep{petri_multi-wavelength_2024-3, petri_multi-wavelength_2026-1}. 
The particle injection scheme plays a central role in controlling the plasma regime and the subsequent radiative properties of the magnetosphere. Moreover, the pair creation process is expected to be highly non-uniform owing to variations in magnetic field curvature, local accelerating electric field and magnetic topology. Such inhomogeneities may translate into inhomogeneous plasma loading of the striped wind and consequently into more complex pulse profiles than simple double peaked shapes. The magnetospheric current density above the polar caps is known to be inhomogeneous, showing in the simplest configuration a dipolar pattern that varies with the magnetic moment inclination $\alpha$ \citep{timokhin_current_2013, gralla_inclined_2017, chernoglazov_coherence_2024}.

In this work, we extend the striped wind framework by introducing spatially dependent emissivity prescriptions that explicitly break high symmetries in the emitting current sheet. We construct a comprehensive atlas of $\gamma$-ray light curves based on the split monopole current sheet geometry, and systematically explore the resulting pulse morphologies. We demonstrate that inhomogeneous emissivity can reproduce a wide variety of observed $\gamma$-ray light curves, thereby reinforcing the current sheet scenario as a robust and physically motivated framework for high-energy pulsar emission.
Our central question in this work is : which parameters actually govern the morphology of $\gamma$-ray light curves, and to what extent does this morphology reflect the geometry and current distribution of the magnetosphere related to pair plasma cascades in the polar caps?
What can the morphology of $\gamma$-ray pulsar light curves tell us about the underlying magnetospheric and geometrical parameters?

The paper is organised as follows. In Sect.~\ref{sec:Modele}, we present the analytical description of the current sheet geometry and detail the plasma loading prescription used to model the emissivity. In Sect.~\ref{sec:atlas}, we compute and analyse the resulting atlas of $\gamma$-ray light curves, emphasising the diversity of pulse morphologies. In Sect.~\ref{sec:Results} we discuss the implications of our results for the interpretation of Fermi-LAT observations and for current-sheet emission models. Finally, in Sect.~\ref{sec:conclusion}, we summarise our main conclusions and outline possible directions for future work.


\section{Analytical model\label{sec:Modele}}


The force-free regime can be regarded as an extreme limit of a pair-filled pulsar magnetosphere. It provides a useful first-order proxy for the expected properties of pulsar radiation signatures. We therefore build high-energy emission models relying on the force-free split monopole solution. The location of the magnetospheric current density and the plasma-loading process are central to the present model, and discussed in the following sections. Dipolar magnetosphere numerical solutions also exist but extracting the radiative signature is much more expensive than for the split monopole for which analytical solutions exist and are used in this work. A detailed treatment of the force-free dipolar radiative wind remains deferred to future work as this requires very intensive numerical computations to construct millions of light curves.

\subsection{Current sheet location}

The most central feature in the $\gamma$-ray emission resides in the location of the emitting current sheet. We identify this region as being the one where the radial component of the magnetic field reverse sign. Several previous works showed that even this surface in the case of a force-free dipole is located nearly at the location of the split-monopole current sheet surface, justifying our approach. The stellar rotation at a rate $\Omega$ along the $\ez$ axis and the radial propagation of the current sheet at a speed $V \lesssim c$ leads to a phase dependence given by $\psi = \varphi - \Omega \, ( t - r /V)$. We use spherical coordinate $(r, \theta , \varphi)$ and $\ez$ is the unit vector along the $z$-axis. The current sheet is then defined by the locus
\begin{equation}
	\cos \theta \, \cos \,\alpha + \sin \theta \sin \,\alpha \cos \psi = 0
\end{equation}
where $\alpha$ is the magnetic obliquity.  This surface is wobbling around the rotation axis and produces the high energy photons in a thin current layer of thickness $\Delta$. The precise radiation mechanism is not specified in the present study and not relevant at this stage of our investigations. More important is the plasma loading process discussed in the next paragraph.

\subsection{Plasma supply}

We model the high-energy emitting region of the striped pulsar wind through a spatially dependent plasma loading prescription that controls how particle density and emissivity are distributed within the current sheet. This approach provides a phenomenological but physically motivated description of the fact that pair supply and dissipation may vary across the sheet, and may vary with both latitude~$\theta$ and azimuth~$\varphi$, breaking the spherical symmetry. Allowing the loading to break the north-south and azimuthal symmetries of the emission zone leads to asymmetric pulse profiles, unequal peak intensities, and a broad range of light-curve morphologies comparable to those observed by Fermi-LAT. In this sense, plasma loading provides a simple but effective parametrisation of the degree to which the current sheet is fed by pairs and radiating, thereby linking the microscopic distribution of emitting particles close to the polar cap to the macroscopic pulse profile in $\gamma$-rays outside the light cylinder.

As a simple prescription, we suppose a variation in pair multiplicity leading to an emissivity depict by spherical harmonic functions $Y_{\ell,m}$ where $\ell$ and $m$ are positive integers labelling the different modes. For simplicity, we set $\ell=m$ for the current density pattern above the polar cap. The density of the pairs escaping the polar cap regions are then modulated by a factor $ \kappa \, Y_{m,m}(\theta',\varphi')$ in the magnetic axis coordinate system where $\kappa$ is a parameter controlling the plasma loading and related to the pair multiplicity. This prescription needs to be rotated into the stellar rotation axis coordinate system $(r,\theta,\varphi)$. It corresponds to a rotation of the spherical harmonic functions performed by Wigner rotation matrices in the general case. However, for $\ell=m$, the transformation simplifies drastically, and we get
\begin{subequations}
	\begin{align}
	Y_{m, m}(\theta,\varphi) & = \left(A^2+B^2\right)^{m/2} \, \cos \left( m \arctan (A,B) \right) \\
	A & = \cos \theta \, \sin \chi + \sin \theta \, \cos \varphi \, \cos \chi \\
	B & = \sin \theta \, \sin \varphi \ .
	\end{align}
\end{subequations}
Because $m$ is an integer, the above expression can be rewritten in a polynomial in $A$ and $B$ to speed up computational time. To mimic the propagation of pairs from the polar caps into the current sheet, we replace the variable $\varphi$ by the phase~$\psi$ similarly to the current sheet location.

\begin{figure}[h]
	\centering
	\includegraphics[width=0.7\linewidth]{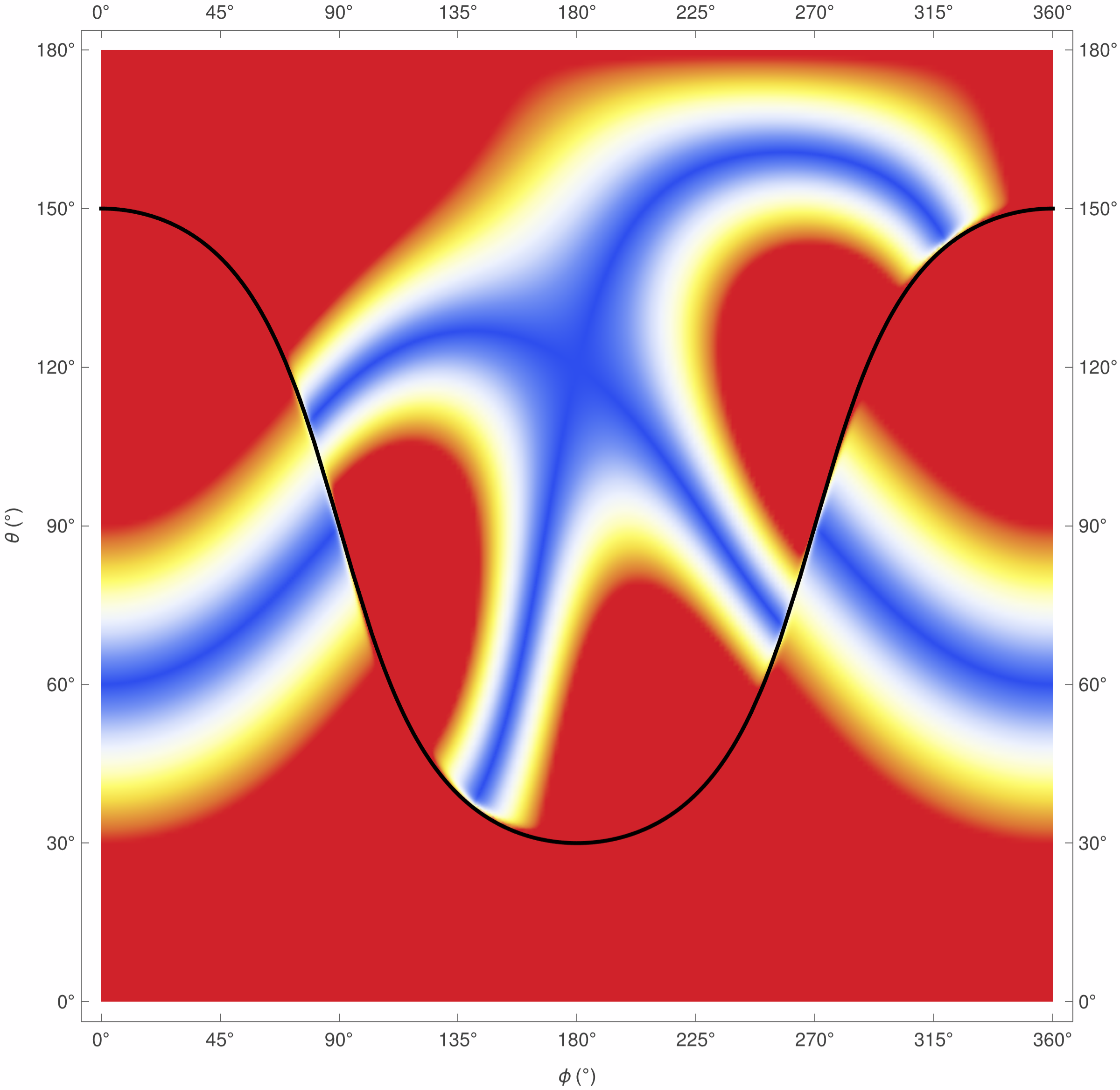}
	\caption{An example of emissivity for an obliquity $\alpha=60\degr$ and the modes $m_{\rm N}=1$ and $m_{\rm S}=2$ in the $(\theta,\varphi)$ plane. The blue colour means zero density whereas red means maximal density normalised to one. The black undulating line depicts the magnetic equator located at $\theta'=\pi/2$.}
	\label{fig:multiplicite_paires_1}
\end{figure}

\begin{figure}[h]
	\centering
	\includegraphics[width=0.7\linewidth]{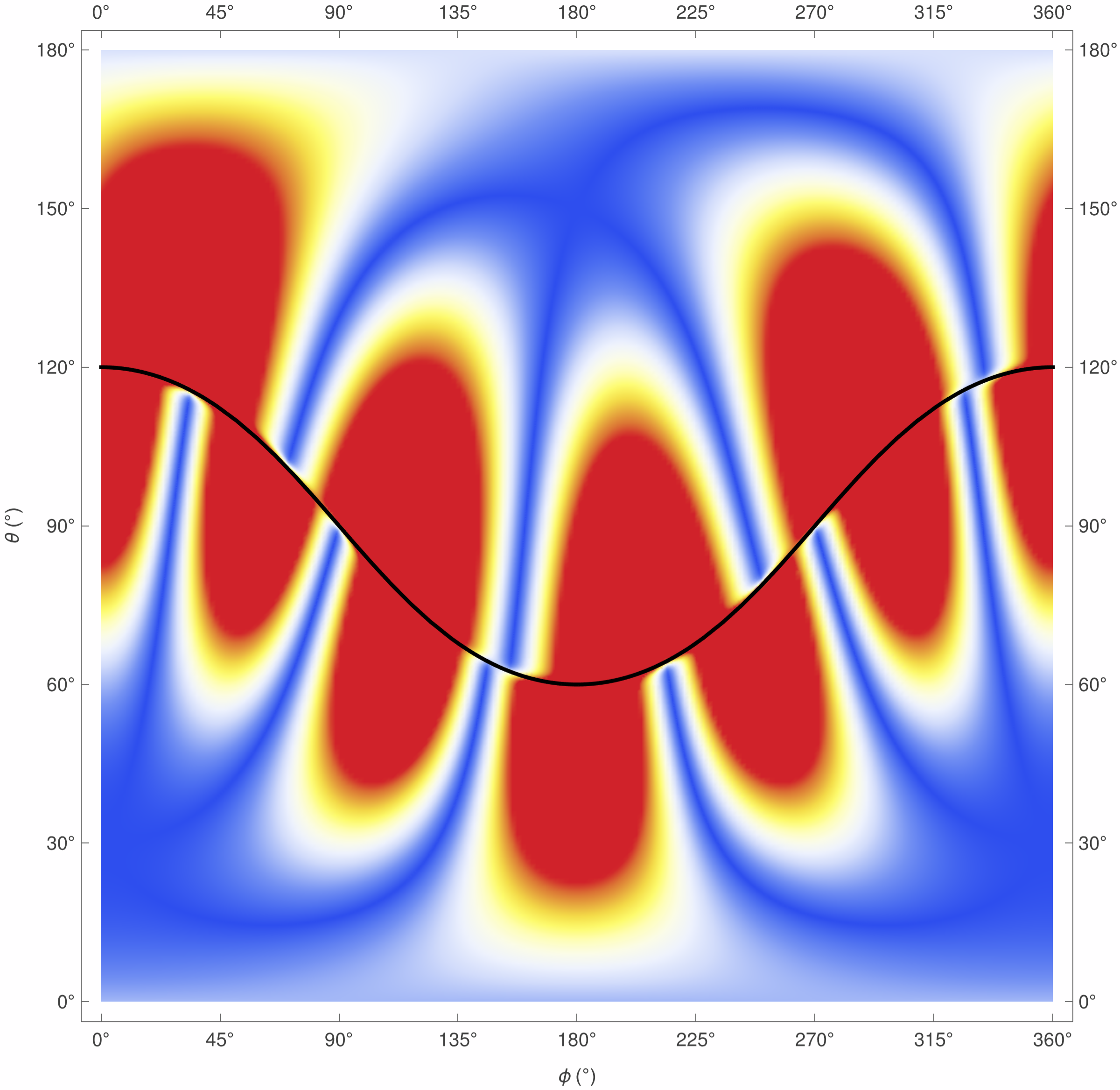}
	\caption{Same as Fig.~\ref{fig:multiplicite_paires_1} but for an obliquity $\alpha=30\degr$ and modes $m_{\rm N}=3$ and $m_{\rm S}=2$.}
	\label{fig:multiplicite_paires_2}
\end{figure}

The key element of our model is a spatially dependent plasma-loading prescription that controls the local pair density and therefore the emissivity~$\epsilon$ within the current sheet. This emissivity function is defined by
\begin{equation}
	\epsilon(t, r, \theta,\varphi) = \epsilon_0 \, f(r, \theta,\varphi; \{\lambda_i\}),
\end{equation}
where $\epsilon_0$ is a normalisation constant and $f$ is a dimensionless function that encodes the spatial dependence of the emissivity. The set of parameters $\{\lambda_i\}$ break the north-south and azimuthal symmetries of the problem. We allow for different emissivities in the North and South Pole, defining two functions $f_{\rm N}$ and $f_{\rm S}$ such that each is written for $K\in[N,S]$ as
\begin{equation}
	f_K(r, \theta,\varphi) = \kappa_K \, | Y^R_{m_K, m_K}(\theta, \varphi, \phi_K) |^2 \, r^{-q} ,
\end{equation}
where we introduced the shifted spherical harmonics $Y^R_{m_K, m_K}(\theta, \varphi, \phi_K)$ and
\begin{itemize}
	\item $\kappa_K$ sets the pair multiplicity in each polar cap,
	\item $m_K$ sets the spherical harmonic number $m$ in each polar cap,
	\item $\phi_K$ introduces a possible rotation of the spherical harmonics around the magnetic axis for polar cap,
	\item $q$ controls the radial decrease of the emissivity.
\end{itemize}
We used squared spherical harmonic functions in order to avoid negative values and cusps when the emissivity vanishes. Fig.~\ref{fig:multiplicite_paires_1} shows a map of the pair plasma supply on the neutron star surface for an obliquity $\alpha=60\degr$ and the modes $m_{\rm N}=1$ and $m_{\rm S}=2$ in the $(\theta,\varphi)$ plane. The blue colour means zero density whereas red means maximal density normalised to one. The black undulating line depicts the magnetic equator located at $\theta'=\pi/2$ (not to be confused with the rotational equator $\theta=\pi/2$).
Fig.~\ref{fig:multiplicite_paires_2} shows a second example of the pair plasma supply on the neutron star surface for an obliquity $\alpha=30\degr$ and the modes $m_{\rm N}=3$ and $m_{\rm S}=2$ in the $(\theta,\varphi)$ plane. Smaller scale structure appear when increasing the azimuthal number $m_K$. Such patterns may plausibly arise from stochastic pair-creation cascades, where plasma columns can form and undergo drift, thereby providing a possible route to the carousel model, originally proposed by \cite{ruderman_theory_1975} to explain drifting sub-pulses.
In the symmetric limit $\kappa_{\rm N} = \kappa_{\rm S}$ and $m_{\rm N} = m_{\rm S} = 0$, we recover the standard case of uniform emissivity within the current sheet, with $\gamma$-ray pulse profiles that were first presented in \cite{petri_unified_2011}.

\subsection{Fitting algorithm}

The light curves, extracted from 3PC, are fitted using the standard $\chi^2$ minimisation with a fixed number of parameters
\begin{equation}\label{eq:chi2}
	\chi^2 = \sum_{i=1}^{N} \left( \frac{y_i-y_{\rm model}}{\sigma_i} \right)^2
\end{equation}
where $N$ is the number of data points, $y_i$ the normalised data points with uncertainty $\sigma_i$ and $y_{\rm model}$ the normalised model. Actually, we give more weight to the peak by modifying this $\chi^2$ into a modified version that we call the score $S_{\!\rm mod}$, expressed as
\begin{equation}\label{eq:score}
	S_{\!\rm mod} = \sum_{i=1}^{N} \left( \frac{y_i-y_{\rm model}}{\sigma_i'} \right)^2 
\end{equation}
where $\sigma_i' = \sigma_i / y_i$ defines an effective weighting scheme that increases the relative contribution of the pulse peaks and reduces the influence of the low-level background emission. This gives greater weight to the stronger peaks during the fitting procedure, rather than to the low-level background emission. Other useful metrics that could be tested include the root mean square error (RMSE) and the mean absolute error (MAE) to assess the robustness of the model ranking. However, we do not consider these alternative metrics because of the substantial computational cost involved.

The complexity of our model is therefore independent of the exact value of the chosen parameters. We emphasise that $(m_{\rm N},m_{\rm S})$ do not represent the number of coefficients in a Fourier expansion of the emissivity pattern but simply two parameters describing one and only harmonics in each polar cap. Consequently, the use of information criteria such as the Bayesian Information Criterion (BIC) or the Akaike Information Criterion (AIC) is not straightforward in the present parametrisation. Nevertheless, producing high harmonics pattern with large values of $m_{\rm N}$ and $m_{\rm S}$ in the pair cascades seems unrealistic. We therefore restrict the modes to $m_{\rm N} \leq 4$ as well as $m_{\rm S} \leq 4$ since increasingly high-order azimuthal structures may be difficult to reconcile with the expected spatial coherence of pair-cascade regions. Moreover, looking for the best fit to each light curve, we could also apply a penalisation depending on chosen mode $(m_{\rm N},m_{\rm S})$ to favour low modes. A general formulation would compute a score according to
\begin{equation}\label{eq:score}
 \textrm{score} = S_{\!\rm mod} + \lambda \, \mathcal{P}(m_{\rm N},m_{\rm S})
\end{equation}
where $\lambda \geq 0 $ is a weight to be adjusted depending on the expression for the penalty $\mathcal{P}(m_{\rm N},m_{\rm S})$. Several choices could be made like $\mathcal{P}(m_{\rm N},m_{\rm S}) = m_{\rm N} + m_{\rm S}$ or $\mathcal{P}(m_{\rm N},m_{\rm S}) = m_{\rm N}^2 + m_{\rm S}^2$ or $\mathcal{P}(m_{\rm N},m_{\rm S}) = \textrm{max}(m_{\rm N}, m_{\rm S})$. We introduced these penalisations that favour lower modes in our fitting algorithm, however, there is no physically motivated reason for such penalisation except to obtain low modes. Thus, we finally decided to use only the chi-squared value $S_{\!\rm mod}$ and check if a posteriori low modes are naturally favoured.

\section{Atlas of gamma-ray light curves\label{sec:atlas}}

Before comparing our model to the observations, let us summarise the different classes of profile that we expect. These classes are not mutually exclusive, as there is a continuous change in the profile shape when the geometric parameters are changing (obliquity, viewing angle, relative phases among others). We also do not pretend to be exhaustive and restrict our discussion to the modes with $m_{\rm N} \leq 1$ and $m_{\rm S} \leq 1$. Actually, most representative light curves are already found with these constraints of low $m_{\rm N}$ and $m_{\rm S}$.

\subsection{Parameter space and sampling}

We explore a multidimensional parameter space defined by the north-south asymmetry amplitude $\kappa_{\mathrm{N,S}}$, the azimuthal phase zero position of both poles, $\phi_{\mathrm{N,S}}$ or equivalently the phase shift between the south and north pole defined by $\Delta \phi = \phi_{\mathrm{S}} - \phi_{\mathrm{N}}$, the magnetic inclination angle $\alpha$, and the observer line of sight inclination angle $\zeta$. This space is sampled using a regular grid for $\kappa_{\mathrm{N,S}}$, $\phi_{\mathrm{N,S}}$, with $11 \times 11 = 121$ points for $\phi_{\mathrm{N,S}}$ and 4~values for the ratio $\kappa_{\mathrm{N}}/\kappa_{\mathrm{S}}$, $(m_{\rm N},m_{\rm S}) \in \{1,2,5,10\}^2$, combined with a set of representative $(\alpha,\zeta)$ pairs. We adopt $35$ inclination angles $\alpha \in \{5^\circ,\dots,175^\circ\}$ and $91$ viewing angles $\zeta \in \{0^\circ,\dots,180^\circ\}$, yielding $11 \times 11 \times 35 \times 91 \times 4 = 1\,541\,540$ distinct geometry configurations for each pair $(m_{\rm N}, m_{\rm S})$ with at least $m_{\rm N} \geq 1$ and $m_{\rm S} \geq 1$. Only $35 \times 91 \times 4 = 12\,740$ distinct geometry configurations for $(m_{\rm N} = 0, m_{\rm S} = 0)$ and $11 \times 35 \times 91 \times 4 = 140\,140$ configurations for $(m_{\rm N} =0, m_{\rm S} \geq 1)$ or $(m_{\rm N} \geq 1, m_{\rm S}=0)$. We computed light curves up to $m_{\rm N} = m_{\rm S} = 4$. Thus, the resulting atlas consists of $16\,549\,260$ light curves. This sampling strategy provides extensive coverage of the emissivity parameter space while maintaining a representative set of geometry configurations for statistical comparisons with observed $\gamma$-ray pulse profiles. 

\subsection{Morphological classification of observed light curves}

A visual inspection of the predicted pulse profiles suggests that most light curves can be grouped into a limited number of recurring morphological families. Although the transitions between these categories are continuous rather than sharply defined, they provide a useful qualitative framework for describing the diversity of the observed emission patterns.

In this paragraph we propose a rough classification of the morphology of light-curves based on the computed light curves, using as a descriptor, the number of peaks, their amplitude ratio, separation, width and bridge emission. We identify the following main morphological classes: single peak symmetric (SPS) with one narrow centred peak, single peak asymmetric (SPA) showing one peak with asymmetry between rising and falling sides, resolved symmetric double peaks (DPS) with two peaks of comparable intensity, separated by a phase $\Delta\phi$, resolved asymmetric double peaks (DPA) with two peaks with different amplitudes, separated by a phase $\Delta\phi$, partially resolved double peaks (DPU) with two symmetric peaks, partially resolved double peaks (DPO) with two peaks with different amplitudes, double peaks with bridge emission (DPB) with two peaks with different amplitude and significant emission between the peaks, separated by a phase $\Delta\phi$, triple peaked profiles, symmetric (TPS) and asymmetric (TPA), with three peaks and multi-peaked profiles (MPX) with four or more peaks, possibly with complex substructure. Table~\ref{tab:morphology_classes} summarises these different classes.
\begin{table*}[h]
	\centering
	\caption{Morphological classification adopted for the pulsar $\gamma$-ray light curves. The classification is based on the number of peaks, their relative amplitudes, widths, phase separation, and the presence of bridge emission.}
	\label{tab:morphology_classes}
	\begin{tabular}{llp{9cm}}
		\toprule
		Class & Abbreviation & Morphological description \\
		\midrule		
		Single peak symmetric &	SPS & Single symmetric peak. \\		
		Single peak asymmetric & SPA & Single asymmetric peak between leading and trailing edges. \\		
		Double peak symmetric & DPS & Two peaks of comparable amplitude without bridge emission. \\		
		Double peak asymmetric & DPA & Two peaks of significantly different amplitudes. \\		
		Double peaks symmetric, partially resolved & DPO & Two unresolved overlapping symmetric peaks. \\		
		Double peaks asymmetric, partially resolved & DPU & Two unresolved overlapping asymmetric peaks. \\		
		Double peaks bridge & DPB & Two peaks with significant bridge emission. \\		
		Triple-peaked symmetric & TPS & Three distinct peaks with same amplitudes and separations. \\		
		Triple-peaked asymmetric & TPA & Three distinct peaks with different amplitudes and separations. \\		
		Multi-peaked & MPX & Four or more distinct peaks and complex substructures. \\		
		\bottomrule
	\end{tabular}
\end{table*}

The various profile morphologies do not form isolated categories but rather define a continuous sequence. At one end lie classical double-peaked profiles with narrow, well-separated maxima and deep inter-peak minima. Increasing bridge emission progressively fills the valley until the two peaks become partially blended and eventually merge into a single broad structure. 
Motivated by this qualitative inspection, we extract a set of quantitative morphological descriptors intended to characterise the pulse profiles in a model-independent manner. These include the number of visible peaks, the number of inferred physical emission components, the phase separation between the dominant peaks, their full widths at half maximum, their relative amplitudes, the bridge intensity, the degree of peak blending, the peak asymmetries, and the normalised background emission level.

\subsection{Representative light curves}

\begin{figure}[h]
	\centering
	\includegraphics[width=0.95\linewidth]{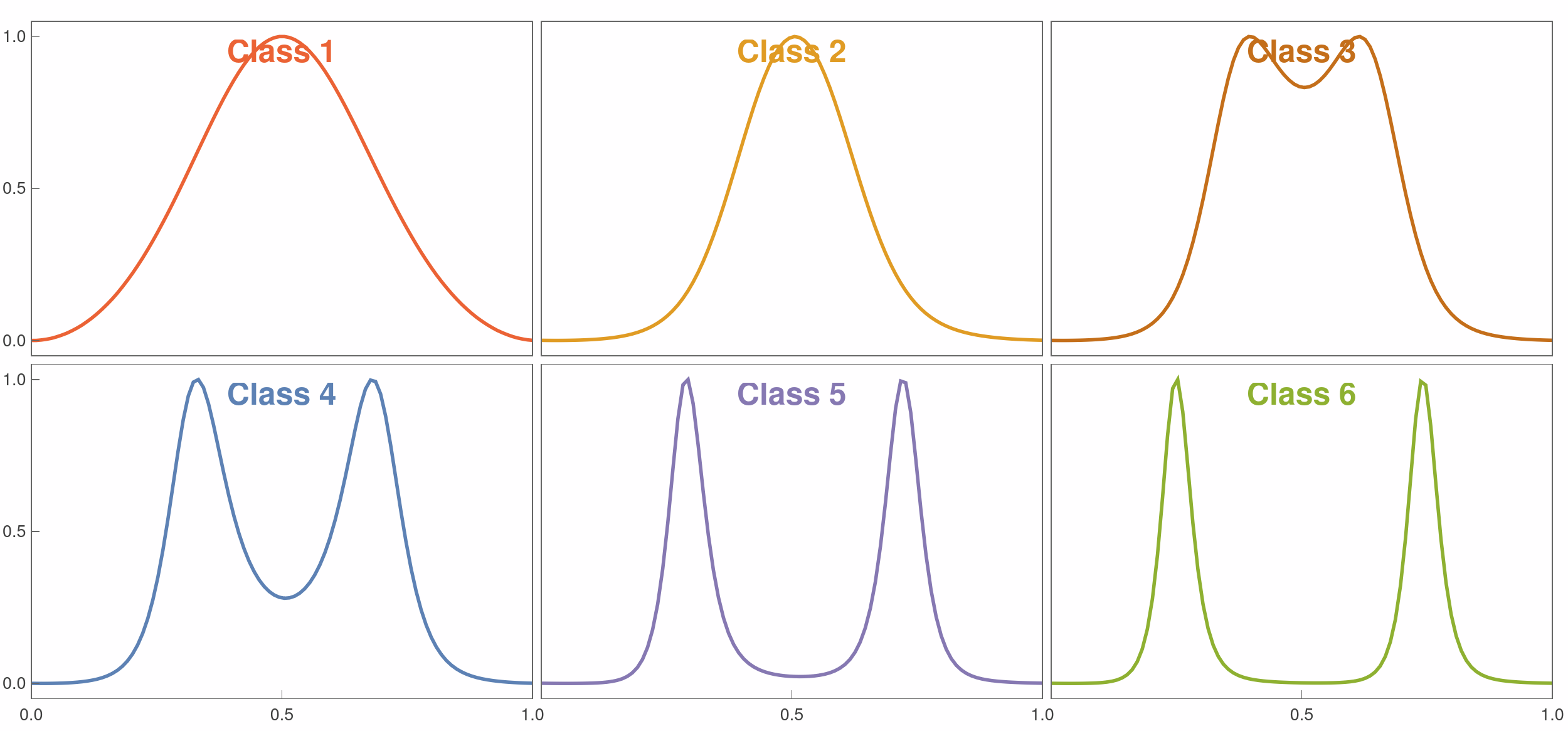}
	\caption{Representative $\gamma$-ray light curves for each morphological class with the parameters $(m_{\rm N}, m_{\rm S}) = (0,0)$ and $\kappa_{\mathrm{S}}/\kappa_{\mathrm{N}}=1$ i.e. a fully symmetric profile. The separation is forced into six classes.}
	\label{fig:representative_lc_0}
\end{figure}

In several previous studies of the split monopole light curve morphology \citep{petri_unified_2011, petri_young_2021} only symmetric profiles were reproduced because of the intrinsic properties of the emitting plasma not showing any anisotropy or inhomogeneity except for a radial spherically symmetric decrease of the density. In the present work, this corresponds to the $(m_{\rm N}, m_{\rm S}) = (0,0)$ mode of pair plasma loading. This configuration can by nature only reproduce symmetric profiles such as SPS, DPS, DPO, see Table~\ref{tab:morphology_classes}. Asymmetric profiles and true bridge emission (not to be confused with overlapping peaks in the symmetric case) are absent in this simple model. Fig.~\ref{fig:representative_lc_0} shows representative light curves classified into six morphological classes for $(m_{\rm N}, m_{\rm S}) = (0,0)$ and $\kappa_{\mathrm{S}}/\kappa_{\mathrm{N}}=1$. The plasma loading is symmetric in this restrictive example.
We check that an asymmetric emissivity with $\kappa_{\mathrm{S}}/\kappa_{\mathrm{N}} \neq 1$ has no significant impact on the peak profile. The change is almost imperceptible and the difference in the off-pulse intensity between both peaks $I^{\rm off}_1$ and $I^{\rm off}_2$ follows approximately the ratio $I^{\rm off}_2 / I^{\rm off}_1 = \kappa_{\mathrm{S}} / \kappa_{\mathrm{N}}$.

\begin{figure}[h]
	\centering
	\includegraphics[width=\linewidth]{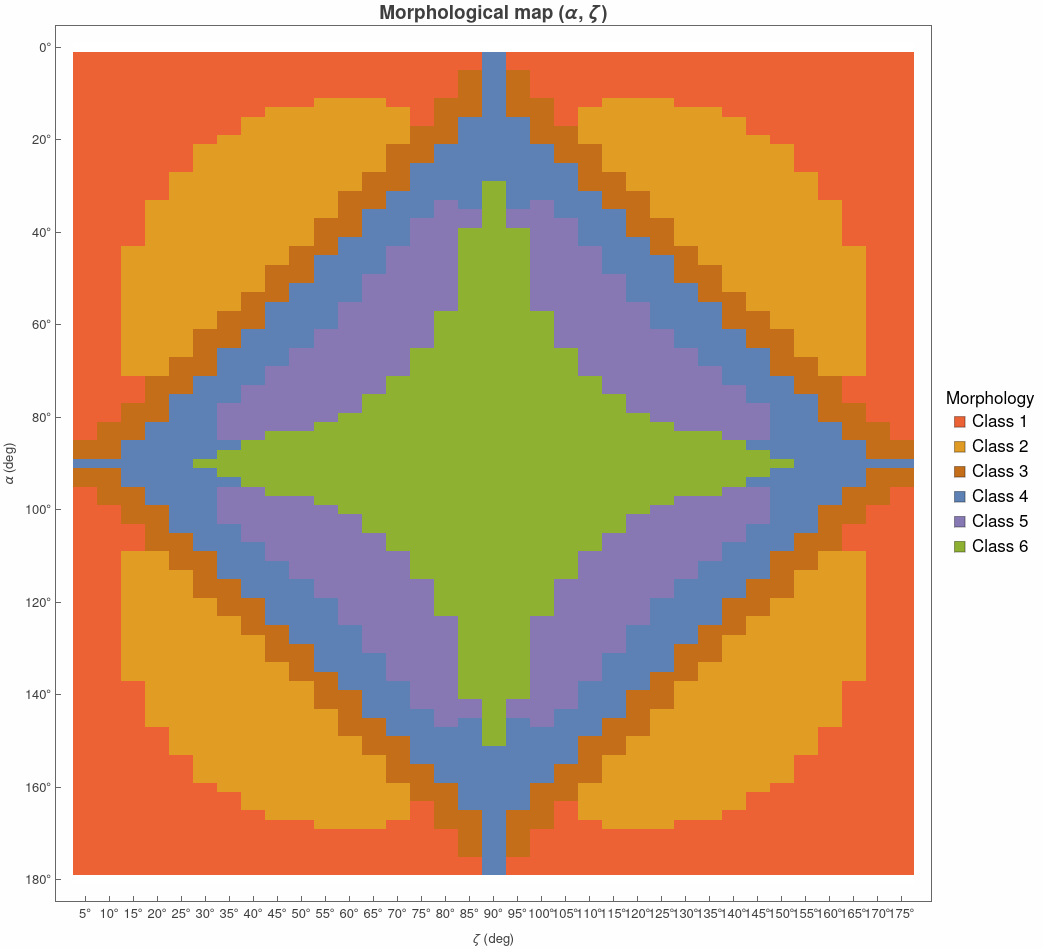}
	\caption{Map of each morphological class in the plane $(\alpha, \zeta)$ for $(m_{\rm N}, m_{\rm S}) = (0,0)$ and $\kappa_{\mathrm{S}}/\kappa_{\mathrm{N}}=1$.}
	\label{fig:representative_lc_atlas}
\end{figure}

A map of each morphological class in the $(\alpha, \zeta)$ plane for $(m_{\rm N}, m_{\rm S}) = (0,0)$ is shown in Fig.\ref{fig:representative_lc_atlas}. Two sharp peaks well separated by almost half a period requires $\alpha \sim 90 \degr$ or $\zeta \sim 90 \degr$. Two overlapping peaks requires $\alpha + \zeta = 90 \degr$ or relations found by symmetry, as is evident from the map. Single peaks are below this line and large single peaks requires $\alpha \sim 0 \degr$ or $\zeta \sim 0 \degr$. This behaviour of the striped wind light curve profile is well known since the work by \cite{petri_unified_2011}.
The inclination $\alpha$ and viewing angle $\zeta$ strongly affect the peak separation and overall shape, as in the symmetric case. 

\begin{figure}[h]
	\centering
	\includegraphics[width=0.95\linewidth]{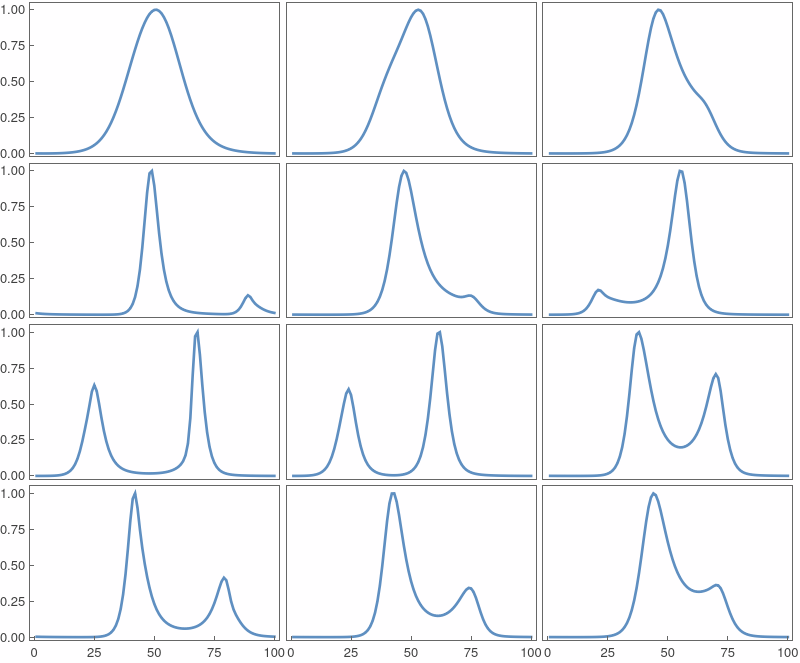}
	\caption{Representative $\gamma$-ray light curves sample for $(m_{\rm N}, m_{\rm S}) = (1,1)$ and $\kappa_{\mathrm{S}}/\kappa_{\mathrm{N}}=1$. The asymmetric nature of these light curves is readily visible.}
	\label{fig:representative_lc_11}
\end{figure}

The situation become more involved when $m_{\rm N}=1$ or $m_{\rm S}=1$ because one or two more degrees of freedom are added, $\phi_{\mathrm{N,S}}$ and $\kappa_{\mathrm{N,S}}$. It is impossible to represent the whole variety of profile. However, a good overview is given by inspecting the morphology for the modes $(m_{\rm N}, m_{\rm S}) = (1,1)$. Figure~\ref{fig:representative_lc_11} shows representative light curves for these morphological classes with $(m_{\rm N}, m_{\rm S}) = (1,1)$. We recover one and two-peak profiles but with clear asymmetry in the peak amplitude and in the leading and trailing part of each pulse.

For $m_{\rm N} \geq 2$ or $m_{\rm S} \geq 2$ the complexity increases with even more morphologies that we not explore in this section. We will show some examples directly applied to the pulsar $\gamma$-ray light curves.

\subsection{Summary of the atlas}

The atlas demonstrates that a large variety of observed $\gamma$-ray light-curve morphologies can be reproduced with only a few free parameters and that asymmetric and inhomogeneous emissivity naturally accounts for asymmetric and multi-peaked profiles without requiring multiple emission zones. The model is particularly efficient in reproducing profiles that are difficult to obtain with symmetric emissivity, producing one two or even three and four peaks depending on the geometry.

\section{Results\label{sec:Results}}

We now describe the results of our fitting procedure applied to a significant fraction of the $\gamma$-ray pulsar catalogue, irrespective of their millisecond or young nature. We apply the split monopole emission model to the most luminous $\gamma$-ray pulsars from the 3PC and allow azimuthal modes up to $m_{\rm N}^{\rm max} = m_{\rm S}^{\rm max} = 4$.

\subsection{Sample}
\begin{figure}[h]
	\centering
	\includegraphics[width=0.95\linewidth]{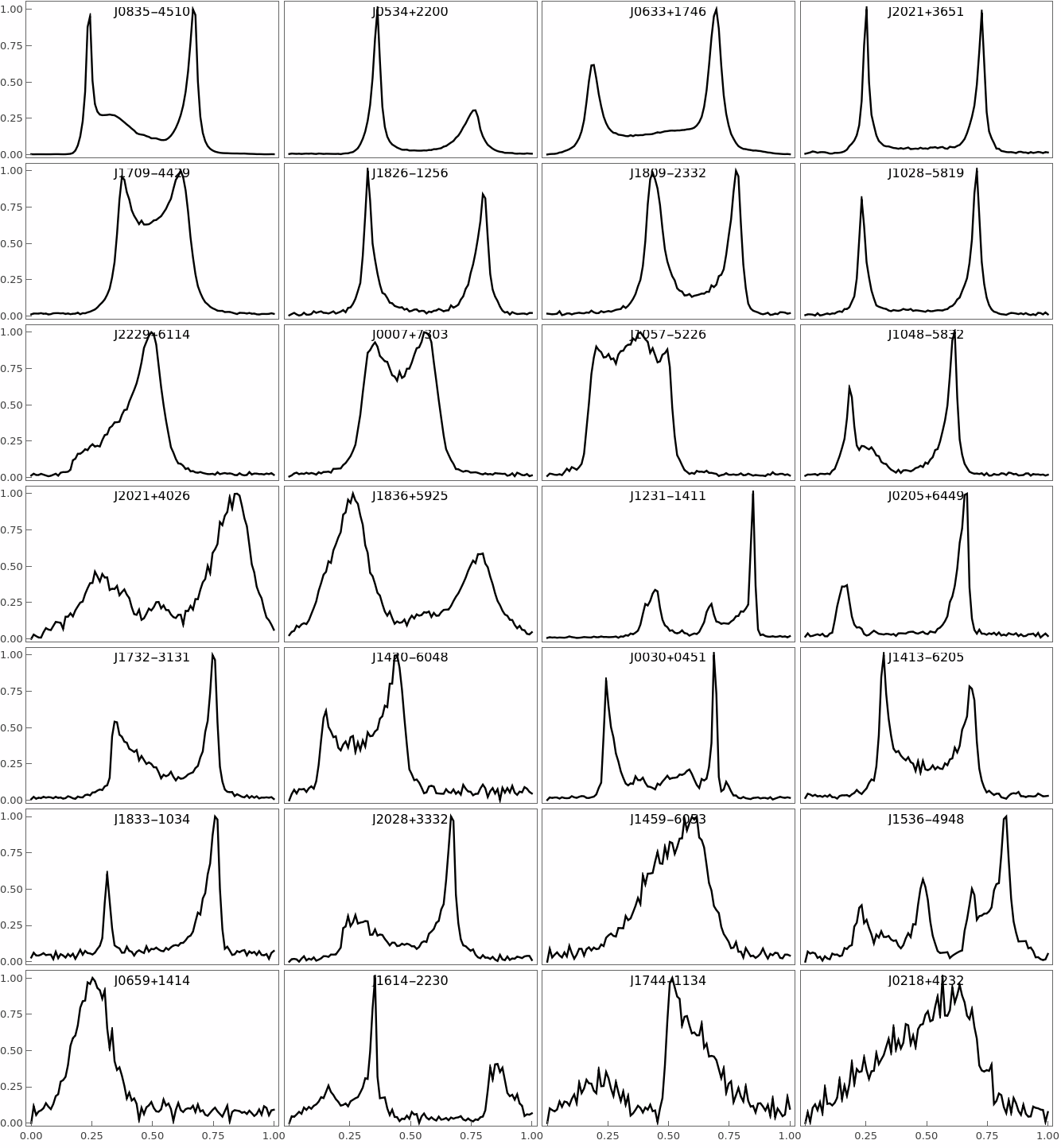}
	\caption{A representative subset of the variety of $\gamma$-ray pulsar light curves extracted from our sample of 130~pulsars.}
	\label{fig:courbes_gamma_echantillon}
\end{figure}

Among the selected 130 pulsars for which their light curves possess more than 100~bins per period, we show representative profiles in Fig.~\ref{fig:courbes_gamma_echantillon}. For those with 200, 400 or 800 bins per period, we re-binned the light curves to 100~bins such that all data have exactly the same number of phase points as the theoretical profiles. This allows us to use straightforwardly the fast Fourier transform for the cross-correlation as explained in \cite{lorange_spin-orbit_2026}. Our sample shows the wealth of profile variety: one symmetric or asymmetric peak, two symmetric or asymmetric peaks, two peak, unresolved or with bridge emission, three peaks and even four peak with complex structure. It includes the Crab (J0534+2200), the Vela (J0835-4510), Geminga (J0633+1746), J1709-4429 and other well known bright $\gamma$-ray pulsars. 


First, we discuss this set of representative pulsars on an individual basis before focusing on a population-level interpretation of our findings. With 130 pulsars in our sample, we can obtain statistically significant results and gain insight into the polar cap plasma loading activity by disregarding the uncertainties in the magnetic structure.

\subsection{Representative pulsars}

The brightest $\gamma$-ray pulsars already illustrate the diversity of observed pulse profiles. We also discuss several pulsars with different morphologies - one peak, two peaks and asymmetric peaks - representative of the full sample, including the most famous ones such as Crab, Geminga and Vela. 

\begin{figure}[h]
	\centering
	\includegraphics[width=\linewidth]{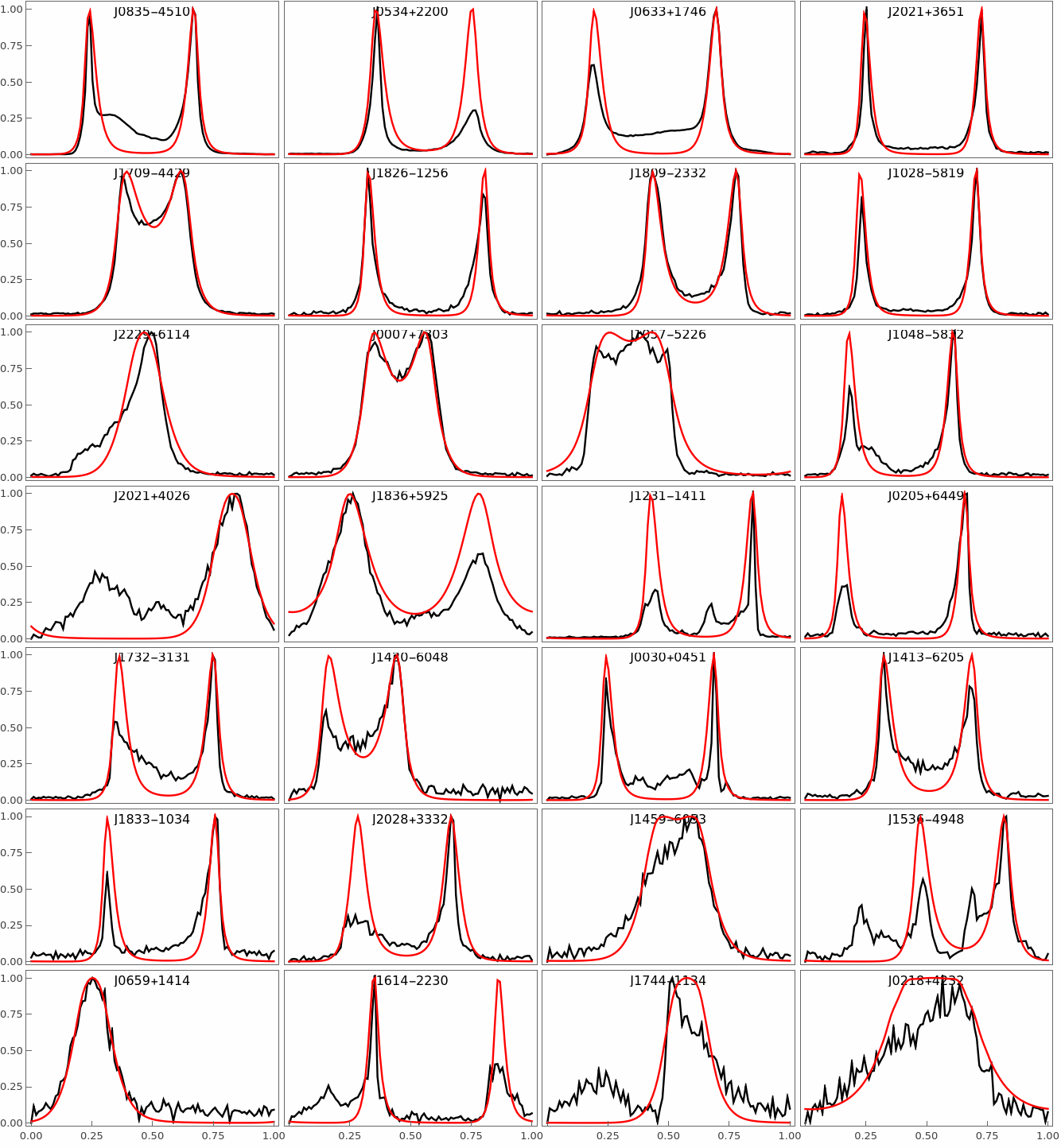}
	\caption{Best model of the pulsars shown in Fig.\ref{fig:courbes_gamma_echantillon} with the symmetric mode ${m_{\rm N}=m_{\rm S}=0}$. Observations are shown in black solid lines and the model in red solid lines.}
	\label{fig:pulsar_echantillon_mn0_ms0}
\end{figure}

\begin{figure}[h]
	\centering
	\includegraphics[width=\linewidth]{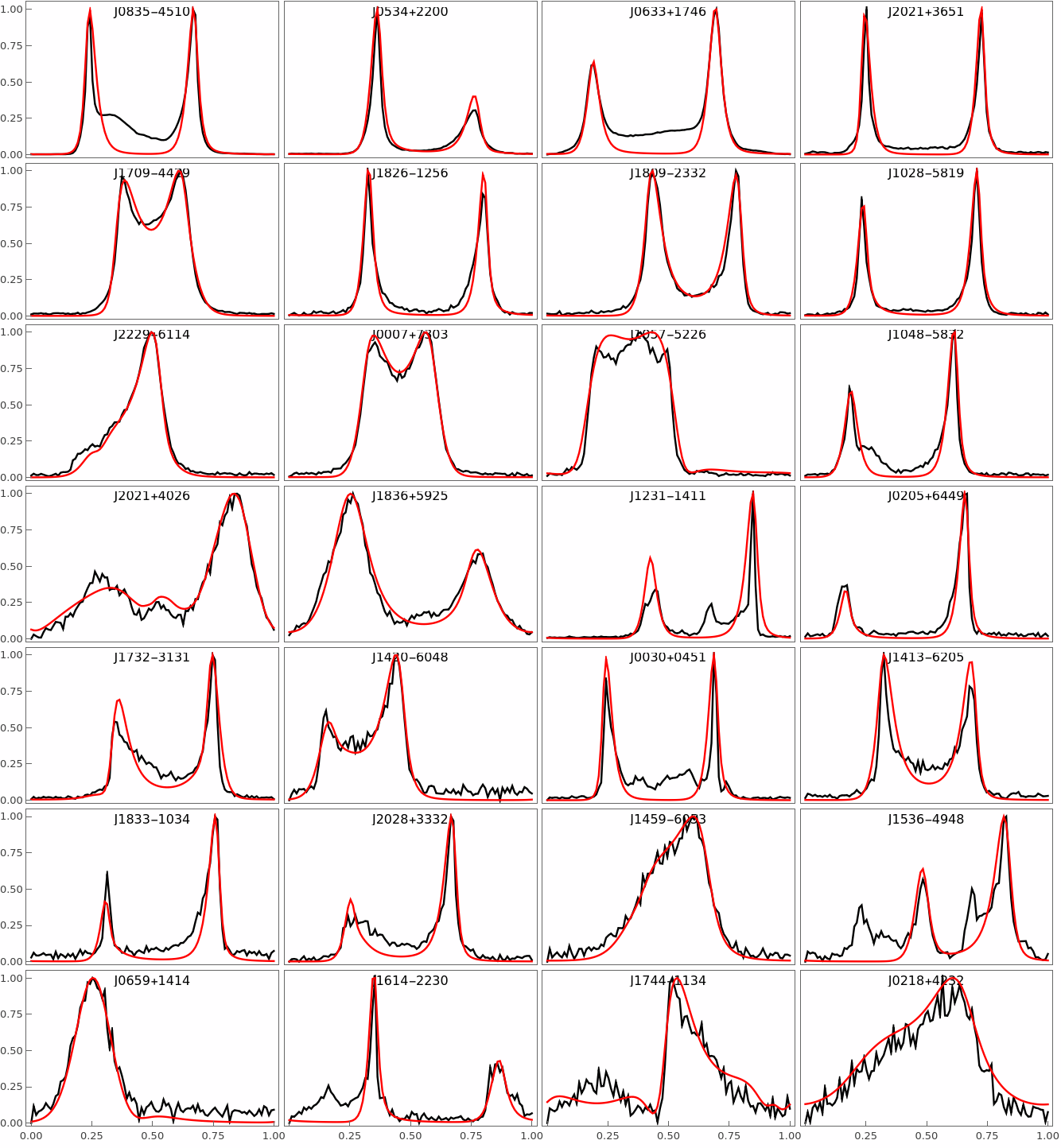}
	\caption{Same as Fig.\ref{fig:pulsar_echantillon_mn0_ms0} but wit the mode $m_{\rm N}=m_{\rm S}=1$.}
	\label{fig:pulsar_echantillon_mn1_ms1}
\end{figure}

\begin{figure}[h]
	\centering
	\includegraphics[width=\linewidth]{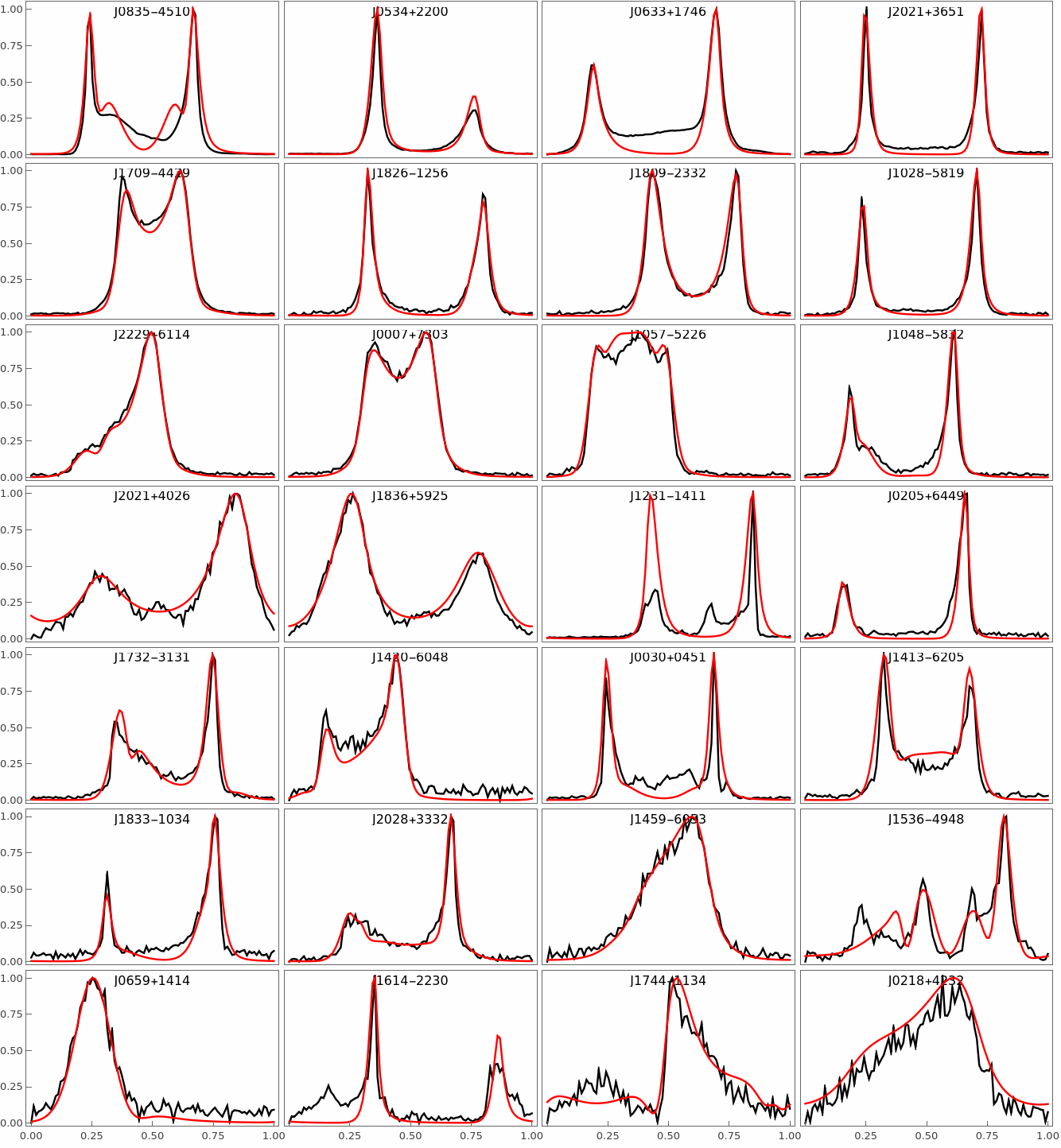}
	\caption{Same as Fig.\ref{fig:pulsar_echantillon_mn0_ms0} but wit the mode satisfying $m_{\rm N} \leq 4$ and $m_{\rm S} \leq 4$.}
	\label{fig:pulsar_echantillon_mnx_msx}
\end{figure}

The best fits for the symmetric mode ${m_{\rm N}=m_{\rm S}=0}$ are shown in Fig.\ref{fig:pulsar_echantillon_mn0_ms0}. Although some pulsars can be explained by this simple model, the vast majority requires at least asymmetric shapes reproduces by the mode $m_{\rm N}=m_{\rm S}=1$ as shown in Fig.\ref{fig:pulsar_echantillon_mn1_ms1}. Finally, Fig.\ref{fig:pulsar_echantillon_mnx_msx} shows the best fit obtained by imposing $m_{\rm N} \leq 3$ and $m_{\rm S} \leq 3$. For some pulsars, the difference between Fig.\ref{fig:pulsar_echantillon_mn1_ms1} and Fig.\ref{fig:pulsar_echantillon_mnx_msx} imperceptible because the mode $m_{\rm N}=m_{\rm S}=1$ is already the best compromise. We will see in the next paragraph discussing on the population-level basis that indeed this $m_{\rm N}=m_{\rm S}=1$ mode is able to reproduce a significant fraction of pulse profiles.

Let us start with the Vela pulsar (J0835-4510), which exhibits two prominent peaks and a weaker third peak in the bridge region. This additional peak makes this pulsar particularly challenging to fit. In contrast, the two main, almost symmetric peaks are separated by approximately half a period and can be modelled using the $m_N = m_S = 0$ symmetric mode. While fitting the third peak is possible, this requires the use of higher modes, such as $(m_{\rm N}, m_{\rm S}) = (3,2)$ for example.
Next the Crab (J0534+2200) shows two asymmetric but well separated peaks that are well approximated by the $m_{\rm N} = m_{\rm S} = 1$ mode. Higher modes could be used to better follow the profile but is unnecessary at this level of our study. 
Geminga (J0633+1746) is well known for its significant bridge emission between two prominent but asymmetric peaks. Although these peaks are easily reproduced by an $m_{\rm N} = m_{\rm S} = 1$ mode, we found it difficult to reproduce the bridge emission, a feature not so common in our atlas except for overlapping peaks.
PSR~J2021+3651 is among the simplest light curve to fit, it has two sharp almost symmetric peaks, well separated by almost half a period. The mode $m_{\rm N} = m_{\rm S} = 0$ is sufficient to fairly reproduce its light curve although here again higher modes would lead to slightly better fitting.
Exactly the same could be said for PSR~J1826-1256. 
PSR~J1709-4429 is an interesting case of unresolved almost symmetric peaks fitted with an $m_{\rm N} = m_{\rm S} = 0$ mode. However, using the $m_{\rm N} = m_{\rm S} = 1$ mode helps to refine the light curve by adding a slight asymmetry in the profile. Finally, the lowest $\chi^2$ is given by a $m_{\rm N} = m_{\rm S} = 2$ mode.
PSR~J0007+7303 possesses a very similar light curve well fitted already with an $m_{\rm N} = m_{\rm S} = 0$ mode.
PSR~J1028-5819 is very like PSR~J2021+3651 but with a slight difference in amplitude between both peaks. Due to this, it requires at least an $m_{\rm N} = m_{\rm S} = 1$ mode to reproduce this asymmetry. 
PSR~J1809-2332 is a typical example of a double peak symmetric, partially resolved profile, with weak bridge emission and well represented by the $m_{\rm N} = m_{\rm S} = 0$ mode. Marginally better results are obtained with the $m_{\rm N} = m_{\rm S} = 1$ mode.
PSR~J2229+6114 is the first example of single asymmetric peak (SPA) that by construction cannot be fitted by the $m_{\rm N} = m_{\rm S} = 0$ mode. It requires at least a $m_{\rm N} = m_{\rm S} = 1$ mode to fairly fit the data. The best fit uses even higher modes making it looking like three overlapping peaks. J1459-6053 resembles PSR~J2229+6114 and can be fitted with the same modes. So is also J0218+4232 with emission almost along the full period.
The profile of PSR~J1057-5226 is particularly difficult to explain with a simple model. Its three blended peaks are hard to reconcile with the $m_{\rm N} = m_{\rm S} = 1$ mode. It is an example of a complex structure with three almost symmetric peaks of similar amplitude (TPS) that can only be explained with at least a $m_{\rm N} \geq 2$ and $m_{\rm S} \geq 2$ mode.
PSR~J1048-5832 resembles the Vela pulsar with two prominent peaks and a third weaker peak in the bridge region. As for Vela, it is hard to fit with low modes but its shape is overall reproduced with an $m_{\rm N} = m_{\rm S} = 1$ mode. The third peak appears with an $(m_{\rm N}, m_{\rm S}) = (4,2)$ mode.
PSR~J2021+4026 is a second example with three asymmetric peaks, two wide peaks separated by a weak narrower peak. It can reasonably be explained by a $m_{\rm N} = m_{\rm S} = 1$ mode although the best model is $(m_{\rm N}, m_{\rm S}) = (3,1)$. 
Similarly J1836+5925 shows a light curve with two wide peaks and a third weak peak maybe interpreted as bridge emission. It can in any case be reproduced with another $m_{\rm N} = m_{\rm S} = 1$ mode although the best model is $(m_{\rm N}, m_{\rm S}) = (3,1)$. 
J1231-1411 shows three sharp peaks even with bridge emission, impossible to reproduce with any values of $m_{\rm N}$ and $m_{\rm S}$ modes. 
Even more difficult is J1536-4948 with for peaks and significant bridge emission. The best we could do is to use a $(m_{\rm N}, m_{\rm S}) = (3,1)$ mode, getting four peaks but not all aligned with observations.
Next PSR~J0205+6449, like the Crab, shows two well separated narrow peaks of different amplitude, but contrary to the Crab, both peaks possess the same width. An $m_{\rm N} = m_{\rm S} = 1$ mode fits well the data. 
J1732-3131 resembles Vela and J1048-5832 and requires a $(m_{\rm N}, m_{\rm S}) = (3,1)$ mode.
J1420-6048 is another example with two overlapping and asymmetric peaks, showing significant bridge emission due to the overlap. We found an $m_{\rm N} = m_{\rm S} = 1$ mode sufficient although higher order modes better fit the most prominent feature. 
For J0030+0451, the two prominent peaks are reproduced in any case, symmetric or asymmetric, but the low level bridge emission is not easily recovered. 
J1413-6205 possesses two almost symmetric peaks with bridge emission. Here also, the two prominent peaks are reproduced in any case, symmetric or asymmetric, but the high level bridge emission is not easily recovered. It requires an $m_{\rm N} = m_{\rm S} = 2$ mode.
J0614-3329 is a typical example of symmetric single peak easily recovered by the symmetric mode $m_{\rm N} = m_{\rm S} = 0$.
J2028+3332 has two asymmetric peaks with significant bridge emission reproduced by a $(m_{\rm N}, m_{\rm S}) = (3,2)$ mode.
J1614-2230 has triple peaks and is not possible to fit all three peaks simultaneously.
J1744-1134 is rather unusual, showing two asymmetric peaks with a sharp leading tail, but can be adjusted by an $m_{\rm N} = m_{\rm S} = 1$ mode.

\subsection{Population-level discussion}

The full sample of light curves with their fits is given in Appendix~\ref{app:full_sample}. There, figure~\ref{fig:meilleur_ajustement_mn1_ms1_q5_asymx_echantillon} shows a comparison between observed light curves and representative models with different couples of $(m_{\rm N} \leq 1, m_{\rm S} \leq 1)$ values: $m_{\rm N}=m_{\rm S}=0$ in dashed blue lines and $m_{\rm N}=m_{\rm S}=1$ in red solid lines. This plot highlights which pulsars are well-fitted by the symmetric model (when the blue and red lines overlap) and which require some asymmetry level. In the same vain, figure~\ref{fig:meilleur_ajustement_mnx_msx_q5_asymx_echantillon} shows in red solid line the best fit by imposing a couple $(m_{\rm N} \leq 4, m_{\rm S} \leq 4)$. The atlas shows that the model reproduces reasonably well the main morphological classes observed in the 3PC catalogue: single and double peaks, asymmetric profiles, bridge emission, and some multi-peaked structures. Our results suggest that asymmetric and inhomogeneous plasma loading may contribute to the observed diversity in the population although asymmetric single and double peaks are easier to reproduce than complex multi-peak emission or some bridge emission. The obliquity $\alpha$ and viewing angle $\zeta$ still play a major role in determining the peak separation and shape but the plasma loading parameters $\kappa_{\mathrm{N,S}}$, $\phi_{\mathrm{N,S}}$ and $\Delta \phi$ control the asymmetry and complexity of the profile going up to four peak for instance for PSR~J1536-4948.

\begin{figure}[h]
	\centering
	\includegraphics[width=0.95\linewidth]{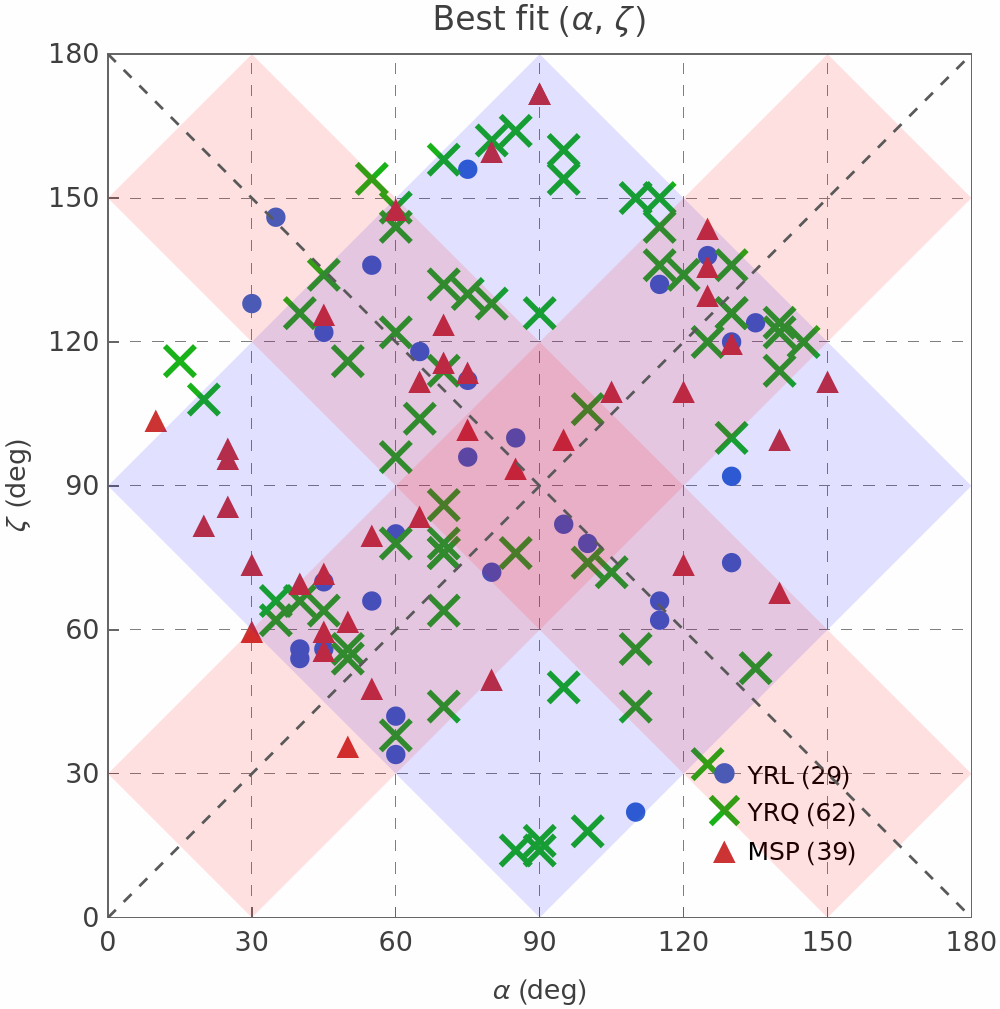}
	\caption{Obliquity $\alpha$ and line of sight angle $\zeta$ found from the best $\gamma$-ray fits. The young radio-loud pulsars, drawn as blue disk (YRL) are distinguished from the young radio-quiet pulsars, drawn as green cross (YRQ) and from the millisecond pulsars, drawn as red triangle (MSP).}
	\label{fig:synthese_geometrie_q5}
\end{figure}

Let us dive more quantitatively into the geometric features found from the sample as a whole. 
First, let us inspect the geometric angles $\alpha$ and $\zeta$ as plotted in the map of Fig.~\ref{fig:synthese_geometrie_q5}. The $\gamma$-ray detectability is shown as a translucence blue square whereas the radio detectability area is shown as a translucence red cross, assuming a fiducial conal beam opening angle of $\rho=30\degr$. The 3PC catalogue lists 29 young radio-loud pulsars (YRL drawn as blue disk), 62 young radio-quiet pulsars (YRQ drawn as green cross) and 39 millisecond pulsars (MSP drawn as red triangle). The frontier between radio-loud and radio-quiet is permeable as some pulsars classified as YRQ by 3PC start to be detected in new survey such as FAST as radio-loud. The latest example in date is PSR~J2238+5903 as reported by \cite{zhang_fast_2026}. Therefore, pulsars categorised as YRQ and located within the red cross area should not be considered as definitive evidence against our model. In the following paragraphs we therefore discard the distinction between radio-loud and radio quiet and only retain the young and MSP classes as distinctive.

\begin{figure}[h]
	\centering
	\includegraphics[width=\linewidth]{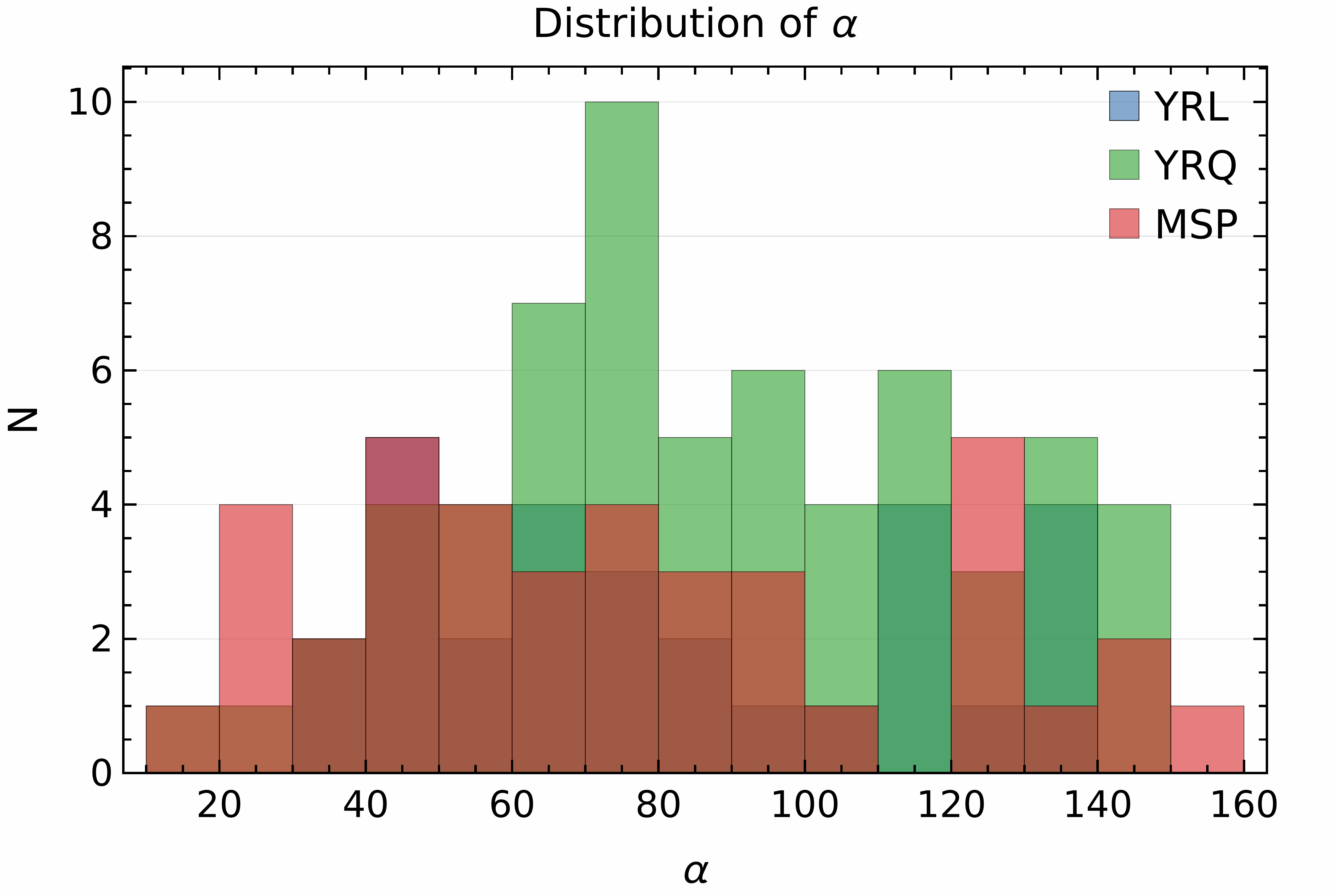}
	\caption{Distribution of the obliquity angle~$\alpha$ among the young radio-loud pulsars (YRL), in blue, young radio-quiet pulsars (YRQ), in green and millisecond pulsars (MSP), in red.}
	\label{fig:histo_alpha}
\end{figure}

\begin{figure}[h]
	\centering
	\includegraphics[width=\linewidth]{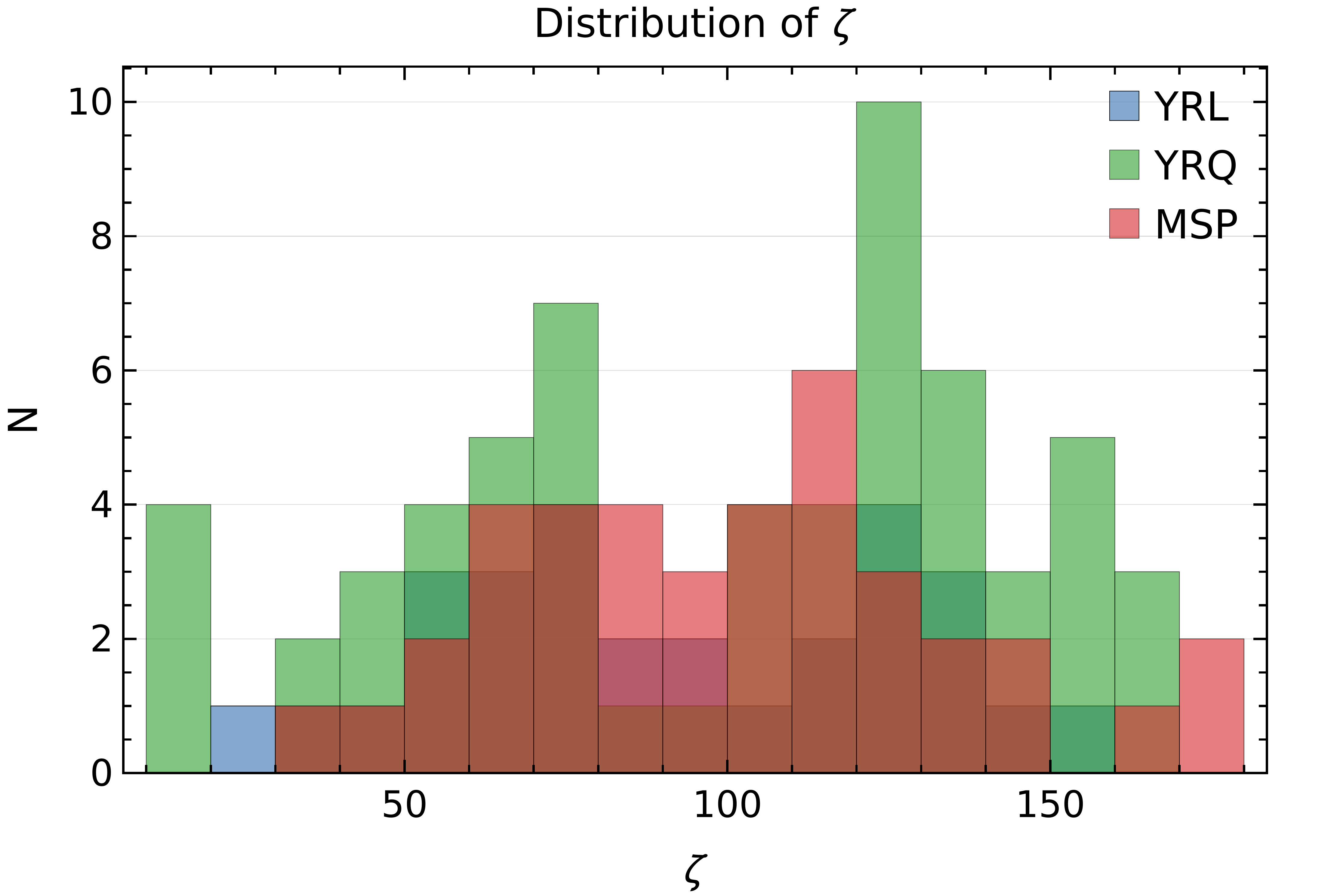}
	\caption{Same as Fig.~\ref{fig:histo_alpha} but for the line of sight viewing angle~$\zeta$.}
	\label{fig:histo_zeta}
\end{figure}

Fig.~\ref{fig:histo_alpha} and Fig.~\ref{fig:histo_zeta} show a synthesis of the best geometric angles $\alpha$ and $\zeta$ respectively. Apart from a trend to accumulate a magnetic orientation towards the rotational equator with $\alpha \sim 90\degr$, we do not observe any significant correlation between the pulsar class and the geometric angles $\alpha$ and $\zeta$.  
This is expected from a geometric point of view, because there is a higher probability to detected pulsed $\gamma$-ray emission for higher obliquity and for line of sight tending to the equatorial plane. 

\begin{figure}[h]
	\centering
	\includegraphics[width=\linewidth]{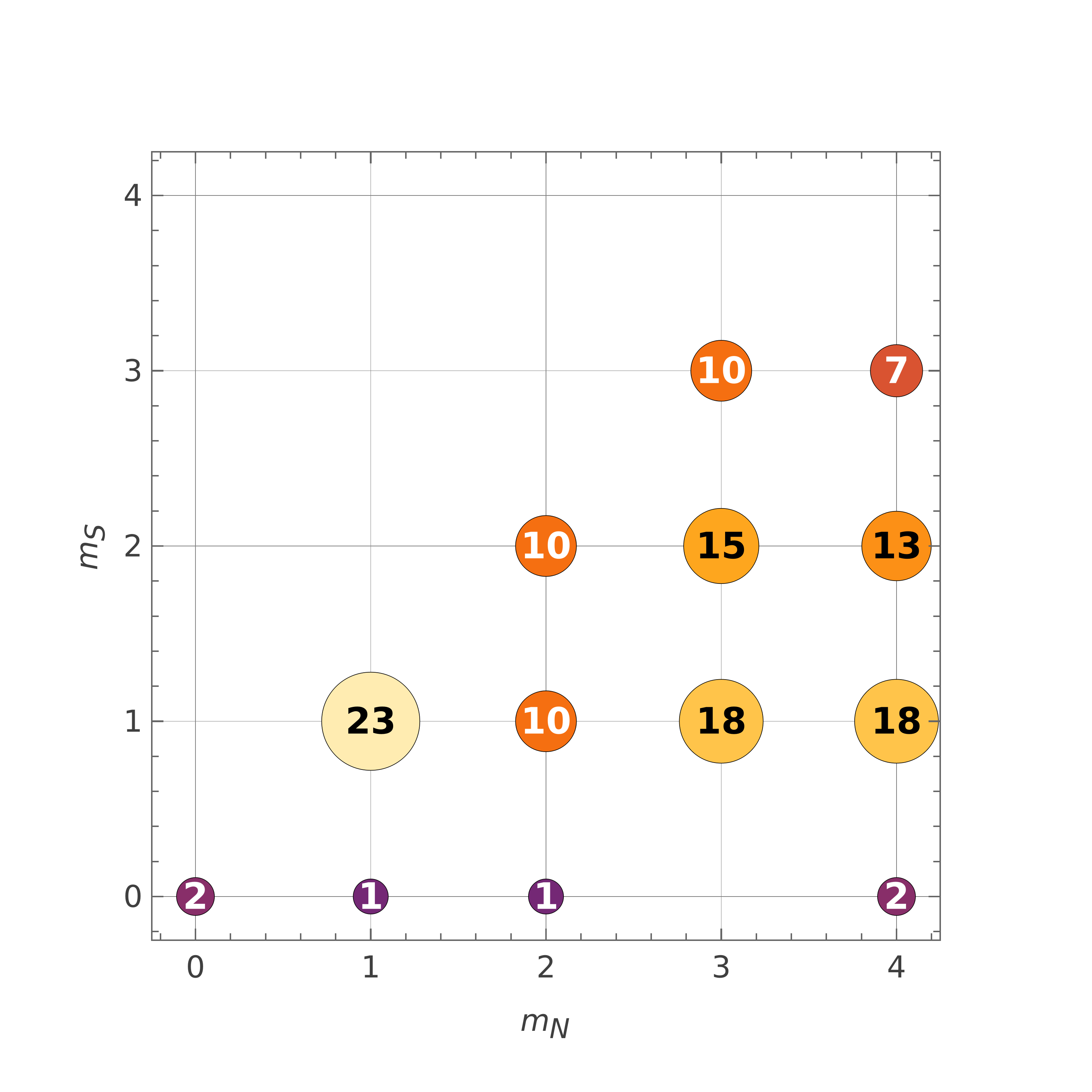}
	\caption{Distribution of the emissivity modulation complexity required to reproduce the observed $\gamma$-ray light curves of the full sample irrespective of the young or MSP nature, summarised by the north and south polar cap modes $m_{\rm N}$ and $m_{\rm S}$, imposing also $m_{\rm S} \leq m_{\rm N}$ due to the north-south symmetry.}
	\label{fig:synthese_modes_q5}
\end{figure}

\begin{figure}
	\centering
	\includegraphics[width=0.95\linewidth]{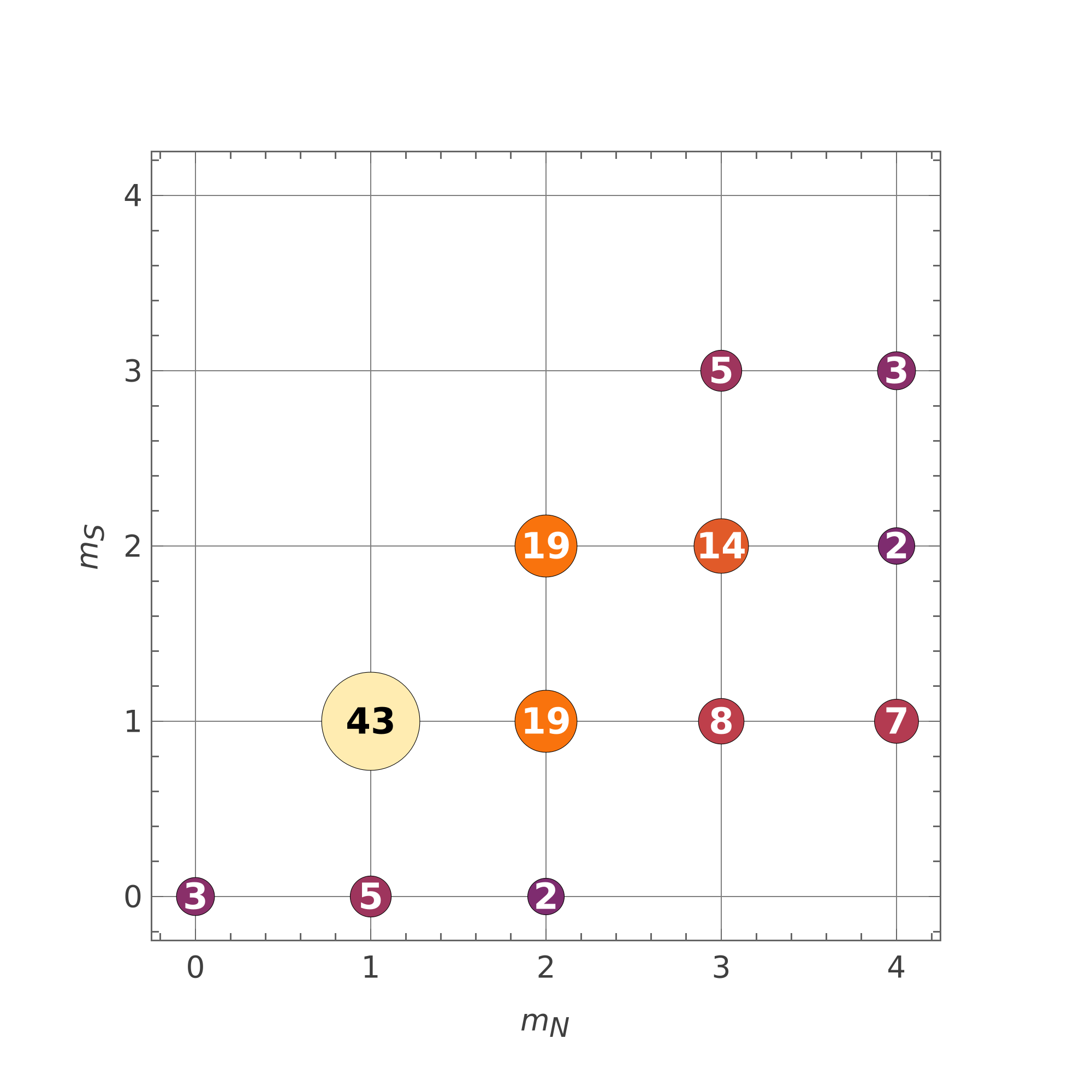}
	\caption{Same as Fig.~\ref{fig:synthese_modes_q5} but choosing the lowest $m$ modes. The $m_{\rm N} = m_{\rm S} = 1$ is clearly dominant.}
	\label{fig:synthese_mnxmsx_minmax_q5}
\end{figure}

Let us now investigate the parameters attached to the effective emissivity modulation and related to the plasma loading configuration. Looking at the mode couple $(m_{\rm N},m_{\rm S})$ in Fig.~\ref{fig:synthese_modes_q5}, we observe naively that low $(m_{\rm N} \leq 1, m_{\rm S} \leq 1)$ modes only fit a small fraction of the pulsars. However, these modes correspond to the best fit found by minimising the $\chi^2$ value Eq.\eqref{eq:chi2}. Some other solutions exist with slightly larger $\chi^2$ values but with similar goodness of fit and lower $m$ modes. With this broader view of the statistics, only complex profiles with three or more peaks would require high modes with $(m_{\rm N}\geq2,m_{\rm S}\geq2)$, see Fig.~\ref{fig:synthese_mnxmsx_minmax_q5}.

Therefore, to better estimate the number of modes required for explaining the light curves, we extract the top-3 models for each pulsar. Table~\ref{tab:synthese_top3} summarises the ranking of the top~3 best models satisfying $m_{\rm N} = m_{\rm S}$. The $m_{\rm N} = m_{\rm S} = 1$ mode is 78 times in the top-3 ranking, corresponding to more than half of the sample (60\%) and the associated median $\chi^2$ is $\chi^2 \approx 259.3$. At the second position, we found the $m_{\rm N} = m_{\rm S} = 2$ mode that is 49 times in the top-3 ranking, corresponding 38\% of the sample and the associated median $\chi^2$ is almost the same with $\chi^2 \approx 259.8$. The $m_{\rm N} = m_{\rm S} = 3$ mode only appears at third position with 40 pulsars in the top-3 ranking, that is 31\% and the associated median $\chi^2$ is $\chi^2 \approx 284.8$. The symmetric $m_{\rm N} = m_{\rm S} = 0$ mode almost never fits the data, even in the top-3 with only 3 pulsars or 2\% of the sample and an associated median $\chi^2$ of $\chi^2 \approx 391.3$. The $m_{\rm N} = m_{\rm S} = 4$ similarly almost never fits the data with an associated median $\chi^2$ of $\chi^2 \approx 359.8$. We found that the $m_{\rm N} = m_{\rm S} = 1$ mode provides the most generic description, able to fit a significant fraction of the pulsars, without introducing too much complexity in the emissivity structure.
\begin{table}[h]
	\centering
	\begin{tabular}{ccrrr}
		\toprule
		$m_{\rm N}$ & $m_{\rm S}$ & Count & Percentage & Median $\chi^2$ \\
		\midrule
		1 & 1 & 78 & 60 & 259.3 \\
		2 & 2 & 49 & 38 & 259.8 \\
		3 & 3 & 40 & 31 & 284.8 \\
		4 & 4 &  3 &  2 & 359.8 \\
		0 & 0 &  3 &  2 & 391.3 \\
		\bottomrule
	\end{tabular}
	\caption{Ranking of the $m_{\rm N} = m_{\rm S}$ modes in the top-3 ranking. The number of pulsars fitted with a particular $m_{\rm N} = m_{\rm S}$ mode is given in the count column, the corresponding percentage in the next column and finally the median $\chi^2$ value for each model.}
	\label{tab:synthese_top3}
\end{table}

\begin{figure*}[h]
	\centering
	\includegraphics[width=0.95\linewidth]{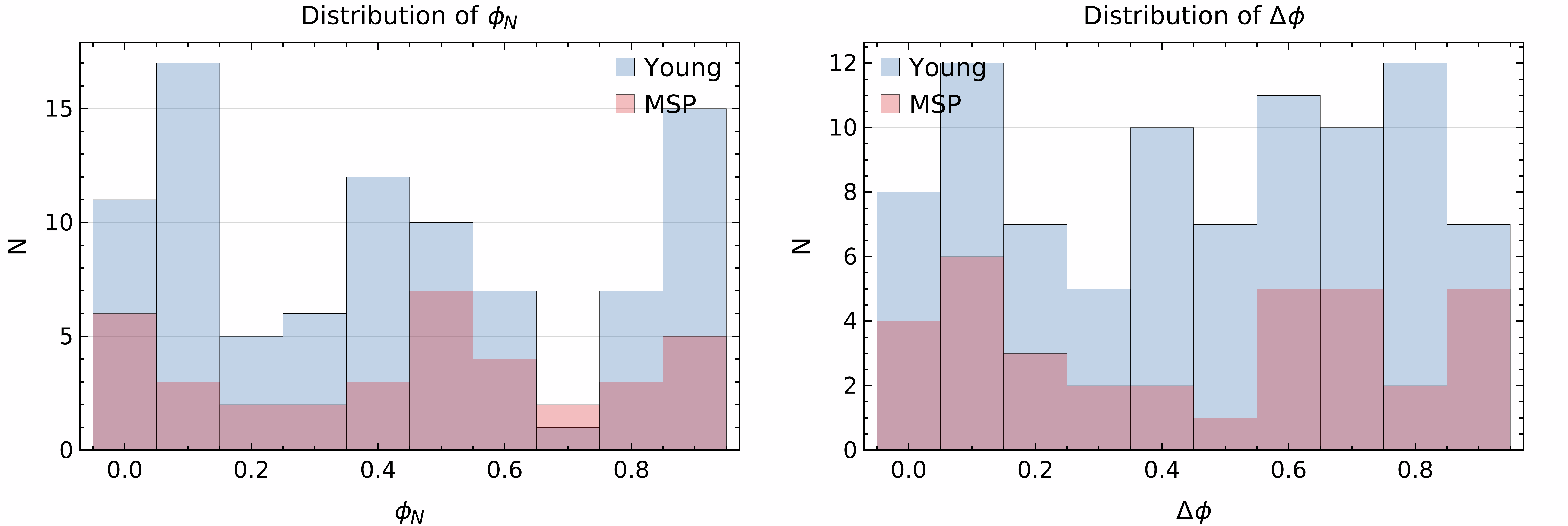}
	\caption{Distribution of the north pole emissivity phase alignment $\phi_{\mathrm{N}}$, on the left histogram, and distribution of the phase shift $\Delta \phi$ between the south and north pole, on the right histogram, for the young pulsars in blue and for the millisecond pulsars in red. Phase alignment values are favoured around  $\phi_{\mathrm{N}}=0$ and $\phi_{\mathrm{N}}=0.5$ for both classes whereas the $\Delta \phi$ distribution is more compatible with a uniform distribution.}
	\label{fig:histo_phi_Dphi}
\end{figure*}

Next the phase shift $\phi_{\rm N}$ and $\Delta \phi$, normalised to the periodicity of the mode $m_{\rm N}$ and $m_{\rm S}$ considered, are plotted in Fig.~\ref{fig:histo_phi_Dphi}, on the left and right panel respectively. The distribution of $\phi$ parameters shows an overshoot around $\phi_{\rm N}=0$ and $\phi_{\rm N}=0.5$. This hints to a peculiar orientation of the emissivity pattern with respect to the plane defined by the rotation and magnetic axes. This configuration could be connected to the electric current density pattern flowing out of the polar caps, as deduced from force-free or resistive simulations. Moreover, because several light curves are best fitted with $m_{\rm S}=0$, the $\Delta\phi$ value is irrelevant and therefore the sample of useful $\Delta\phi$ values is lower than that of $\phi_{\rm N}$. No strong preference for particular values of $\Delta\phi$ is evident. The misalignment in the north and south patterns hints to uncorrelated polar cap geometries.

\begin{figure}[h]
	\centering
	\includegraphics[width=0.95\linewidth]{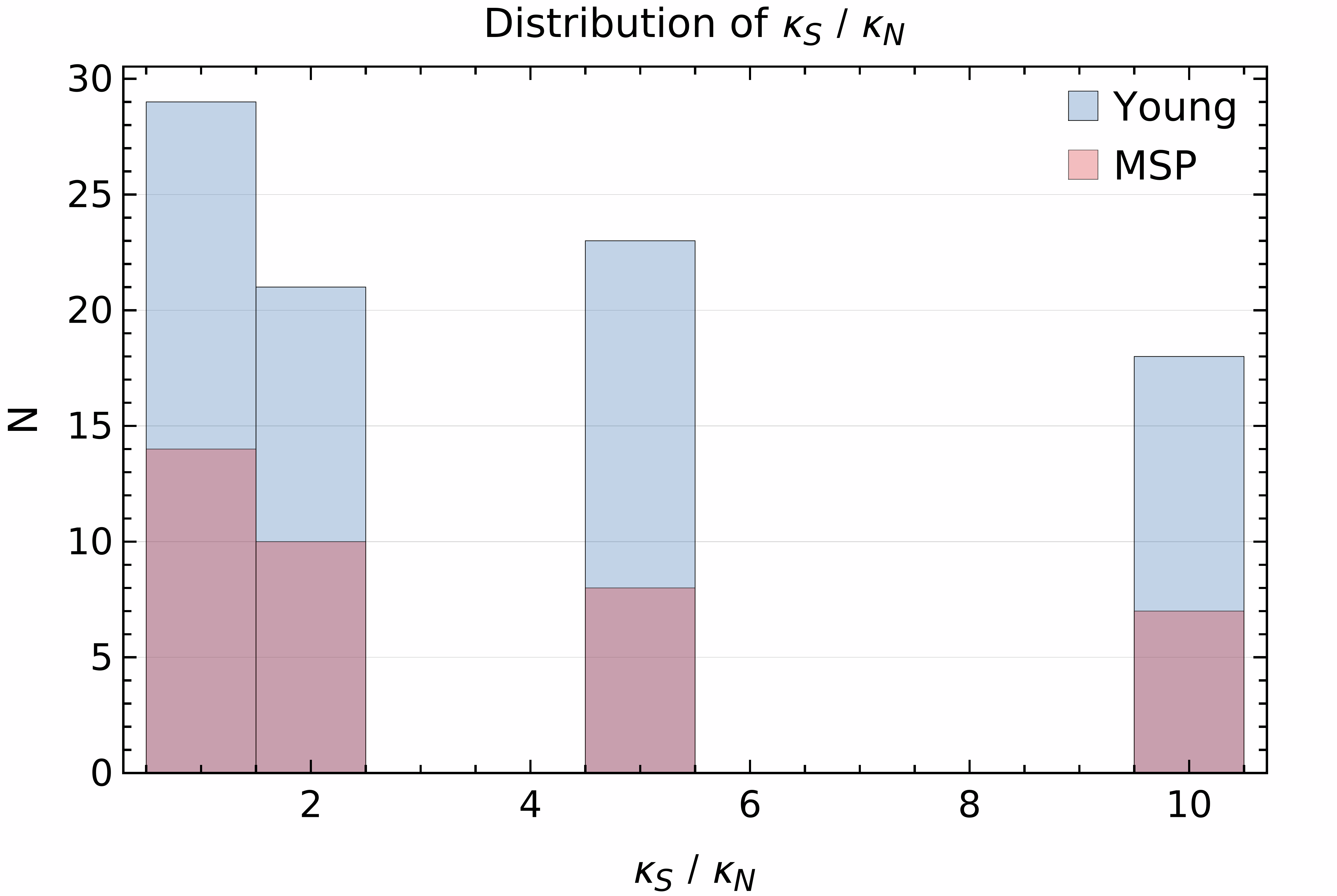}
	\caption{Distribution of emissivity asymmetry $\kappa_{\mathrm{S}}/\kappa_{\mathrm{N}}$ between the south and north pole for the two classes of pulsars: young in blue and MSP in red. The gaps are artifact due to our chosen discrete values being "only" $\kappa_{\mathrm{S}}/\kappa_{\mathrm{N}} = \{1,2,5,10\}$.}
	\label{fig:histo_asym}
\end{figure}

Finally, Fig.~\ref{fig:histo_asym} shows the required asymmetry to reconstruct the light curves. Interestingly, only one fourth of the sample requires symmetric emissivity, many pulsars requiring mildly $\kappa_{\mathrm{S}}/\kappa_{\mathrm{N}}=2$ to high asymmetries $\kappa_{\mathrm{S}}/\kappa_{\mathrm{N}} =5,10$. There is indeed no particular reason to expect a perfectly symmetrical plasma outflow between north and south pole leading to a perfect north-south emissivity. Interestingly also, for symmetric pulse profiles fitted by the $(m_{\rm N},m_{\rm S}) = (0,0)$ mode or by the $(m_{\rm N},m_{\rm S}) = (1,1)$ mode, the light curves are relatively insensitive to the asymmetry parameter, although there are cases where this asymmetry improves the fits. Thus, nearly symmetric pulse profiles are observed more often, even in the presence of strong asymmetries, whereas more complex profiles that are truly sensitive to emissivity asymmetry occur less frequently.

\subsection{Link between radio and $\gamma$-ray observations}

Our emissivity prescription could be related directly or indirectly to the pair plasma loading from the polar cap. However, firm conclusions are difficult to drawn because of several effects perturbing the particle distribution function between their birthplace in the cascade and the current sheet where high energy emission is produced. There exist however avenues to get insight into this connection between the polar caps and the current sheet. For instance, based only on the geometric configuration of the magnetic field $\alpha$ and the orientation of the telescopes with respect to the pulsar rotation axis $\zeta$, we can check for the radio and $\gamma$-ray visibility for a given observer. But particle producing the radio emission are also responsible for high energy emission. Therefore, a direct connection between the polar cap pair creation and the $\gamma$-ray pulse profile could be checked by detecting for instance $\gamma$-ray pulse drifting as in radio pulses. This would establish a firm relation between the radio emitting population and the $\gamma$-ray emission population. 
Also, could a complex radio pulse profile then reflect into a complex $\gamma$-ray light-curve? This second implication would be easier to check than the first expectation because of the low photon statistics at high energy compared to radio wavelength. Some observations have already revealed a correlation between glitches and the $\gamma$-ray luminosity and pulse profile in PSR~J2021+4026, suggesting that the $\gamma$-ray variability may be associated with some electromagnetic activity close to the surface \citep{zhao_mode_2017}.

\subsection{Possible physical interpretation of the emissivity pattern}

The parameters $\kappa_{\mathrm{N,S}}$ and $\phi_{\mathrm{N,S}}$ (or $\Delta\phi$) can be interpreted in terms of physical processes related to pair creation in the polar caps. Indeed, $\kappa_{\mathrm{N}} / \kappa_{\mathrm{S}} \neq 1$ could result from an asymmetric pair production rate between the northern and southern hemispheres, possibly due to asymmetric magnetic field structure (connected for instance to an offset dipole or small scale multipolar structures) and global electric current asymmetries. The phases $\phi_{\mathrm{N,S}}$ could reflect to azimuthal variations in the pair plasma supply, intermittency in pair creation and even drifting of plasma columns within the polar caps, as indirectly seen in drifting radio sub-pulses. Farther away, in the current sheet, localised dissipation regions and reconnection variability could also account for inhomogeneities in the emissivity.

In global simulations, the current sheet is known to be highly structured and intermittent, with localised regions of strong dissipation and particle acceleration. Our phenomenological emissivity prescription captures this behaviour in a simplified way, allowing us to explore its impact on the predicted $\gamma$-ray light curves. Low-order modes with $m \leq 1$ reproduce a large sample of the observed light curves, suggesting possibly that the pair loading modulation may be dominated by large-scale asymmetries rather than small scale structures. We emphasise that the causal link between pair plasma loading and high-energy emission, as the origin of the observed morphological diversity of pulsar light curves, is not yet firmly established. Nevertheless, we are searching for strong observational and theoretical hints, and the present work provides a first step towards elucidating this connection.

More than 90\% of the observed pulsars are reproduced using only dipolar ($m=1$) or quadrupolar ($m=2$) plasma loading modes, indicating that only large scale asymmetries dominate the current-sheet emissivity.
After removing the intrinsic rotational degeneracy of each azimuthal mode, the relative orientation of the northern and southern loading patterns is consistent with a uniform distribution. This suggests that the plasma-loading patterns of the two polar caps are uncorrelated.
%
The reduced azimuthal offset $\Delta\phi$ is defined modulo the intrinsic rotational symmetry of the azimuthal mode and normalised in the interval $[0,1]$. Consequently, $\Delta\phi=0$ and $\Delta\phi=1$ correspond to the same relative orientation, and the apparent accumulation at both ends of the histogram reflects a single preferred alignment rather than two distinct populations.
We found no significant correlation between the parameters of our emission model, neither for the full sample nor for the subset of YRL, YRQ and MSP.

\begin{figure}[h]
	\centering
	\includegraphics[width=\linewidth]{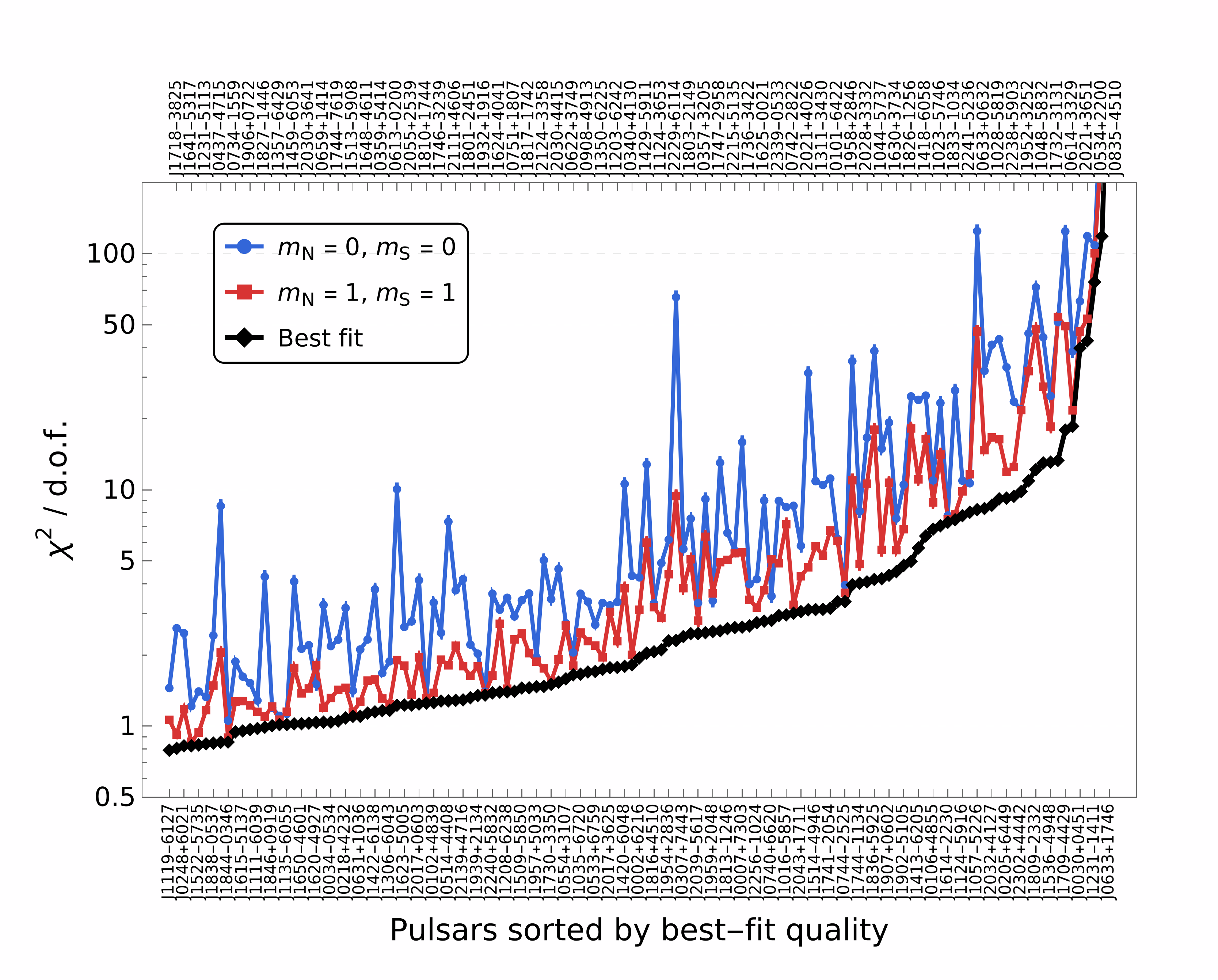}
	\caption{Comparison of the $\chi^2/$d.o.f. values between the symmetric mode $(m_{\rm N},m_{\rm S}) = (0,0)$, in blue, the lowest asymmetric mode $(m_{\rm N},m_{\rm S}) = (1,1)$, in red, and the best fit solution $(m_{\rm N} \leq 4, m_{\rm S} \leq 4)$, in black. Pulsars are sorted by increasing value of the best $\chi^2$ fit. The brightest pulsars (Crab, Vela, Geminga among others) are badly fitted, explaining the high  $\chi^2/$d.o.f. value.}
	\label{fig:chi2_comparison_main}
\end{figure}

Fig.~\ref{fig:chi2_comparison_main} quantifies the decrease in $\chi^2/$d.o.f. value when shifting from the symmetric mode $(m_{\rm N},m_{\rm S}) = (0,0)$, in blue, to the simplest asymmetric mode $(m_{\rm N},m_{\rm S}) = (1,1)$, in red, and finally to the best fit solution $(m_{\rm N} \leq 4, m_{\rm S} \leq 4)$, in black. The brightest pulsars are generally on the right sight of this plot, and are badly fitted because of the small uncertainties in the intensity. This explains their high $\chi^2/$d.o.f. value, although the peaks are reasonably well reproduced. The name of each pulsar is indicated in either the top or the bottom axis.

\begin{figure}[h]
	\centering
	\includegraphics[width=\linewidth]{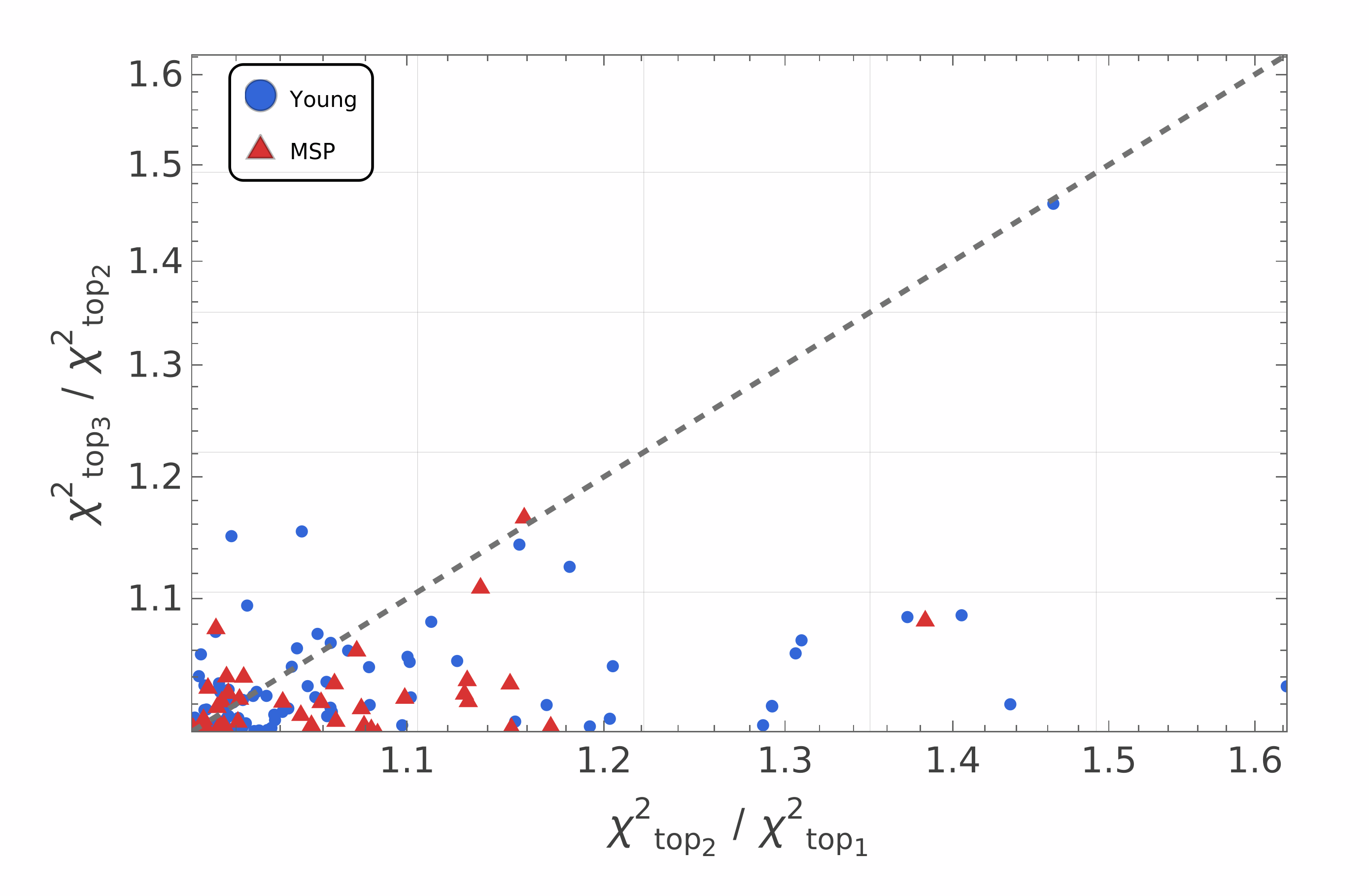}
	\caption{Diagram showing the $\chi^2$ ratios for the top-3 best solutions for each pulsar, on the x-axis $\chi^2_{{\rm top}_2}/\chi^2_{{\rm top}_1}$ and on the y-axis $\chi^2_{{\rm top}_3}/\chi^2_{{\rm top}_2}$. Blue dots correspond to young pulsars and red triangles to MSP. The $\chi^2$ values are all comparable.}
	\label{fig:chi2top_ratios_log}
\end{figure}

Fig.~\ref{fig:chi2top_ratios_log} compares the $\chi^2$ value for the top-3 best models for each pulsar. The x-axis reports the $\chi^2_{{\rm top}_2}/\chi^2_{{\rm top}_1}$ ratio whereas the y-axis reckons the $\chi^2_{{\rm top}_3}/\chi^2_{{\rm top}_2}$ ratio. We distinguish the young pulsars depict as blue dots from the MSP depict as red triangles. The increase in $\chi^2$ is marginal between the three solutions, they are therefore all acceptable for fitting the data according to this criterion.

\begin{figure}[h]
	\centering
	\includegraphics[width=0.9\linewidth]{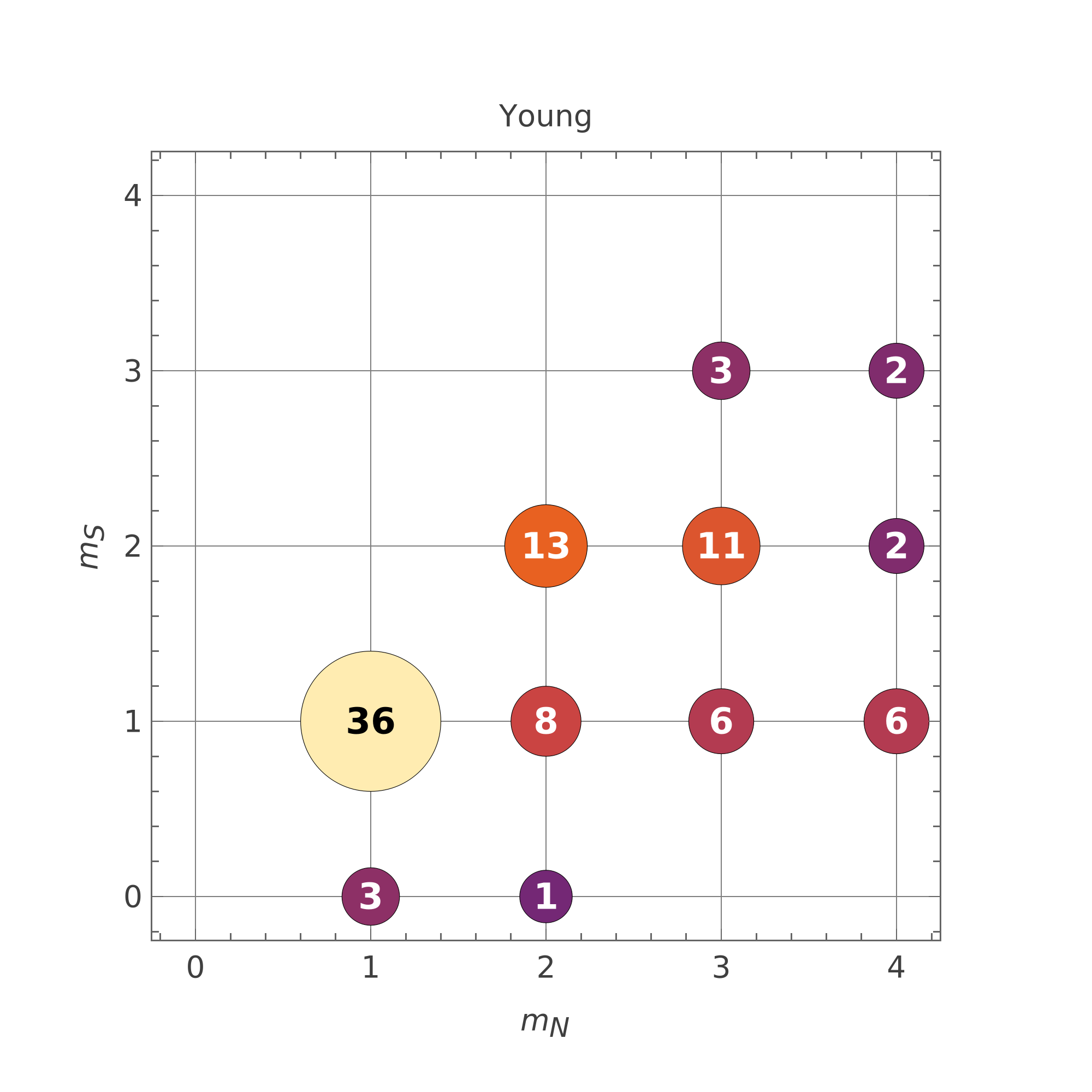} \\
	\includegraphics[width=0.9\linewidth]{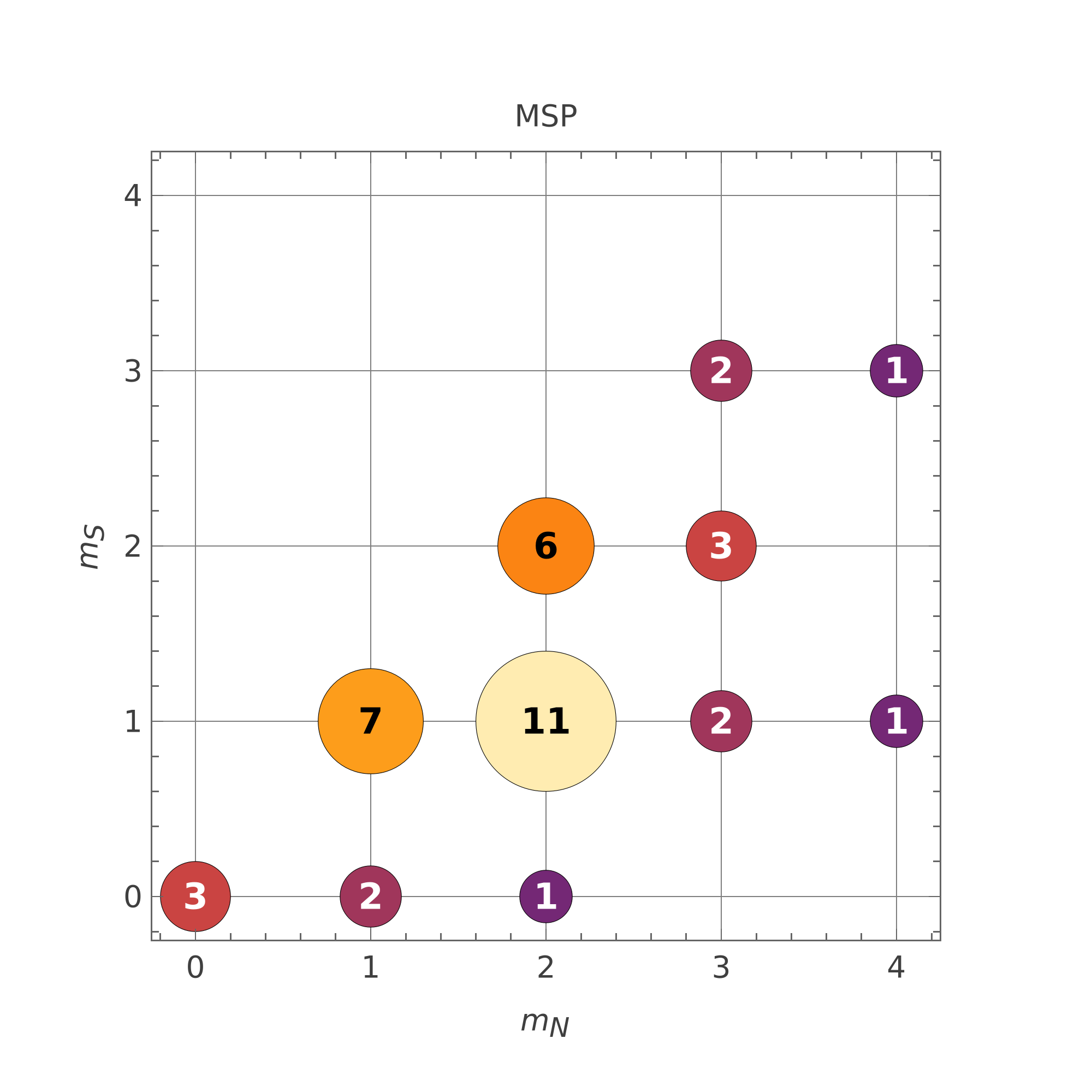}
	\caption{Distribution of the minimal modes $(m_{\rm N}, m_{\rm S})$ required to reasonably fit separately the young pulsars, top panel, and the millisecond pulsars, bottom panel.}
	\label{fig:synthese_mnx_msx_minmax_q5_jeunes_msp}
\end{figure}

The question therefore arises on which model to pick out according to supplementary constraints. One of the main criteria is certainly physical simplicity. Consequently, for parsimony reasons on the modes $(m_{\rm N}, m_{\rm S})$, we choose the modes with the lowest maximum $m_{\rm x} = m_{\rm N}$ or $m_{\rm S}$ as a reasonable solution. Performing this choice of the parsimonious mode from the top-3 list, we get the mode distribution shown in Fig.~\ref{fig:synthese_mnx_msx_minmax_q5_jeunes_msp}. We clearly see that young pulsars mainly require $(m_{\rm N}, m_{\rm S})=(1,1)$ modes whereas MSPs require mainly at least one pole with an $m_{\rm x}=2$ mode. The light-curve fits are statistically better represented by higher-order emissivity patterns in MSPs, which may be consistent with a stronger contribution of non-dipolar magnetic structures, although the strong degeneracy between viewing geometry and plasma-loading parameters prevents a unique physical interpretation. 
Moreover, the mode $(m_{\rm N}, m_{\rm S})=(2,1)$ is the most frequent case for MSP, hinting to highly asymmetric magnetic field structure between the north and south pole. The $\gamma$-ray emission is therefore also able to probe the magnetic field structure at low altitude through the inhomogeneous transport of the radiating particles along the magnetic field lines.

\begin{figure}[h]
	\centering
	\includegraphics[width=0.9\linewidth]{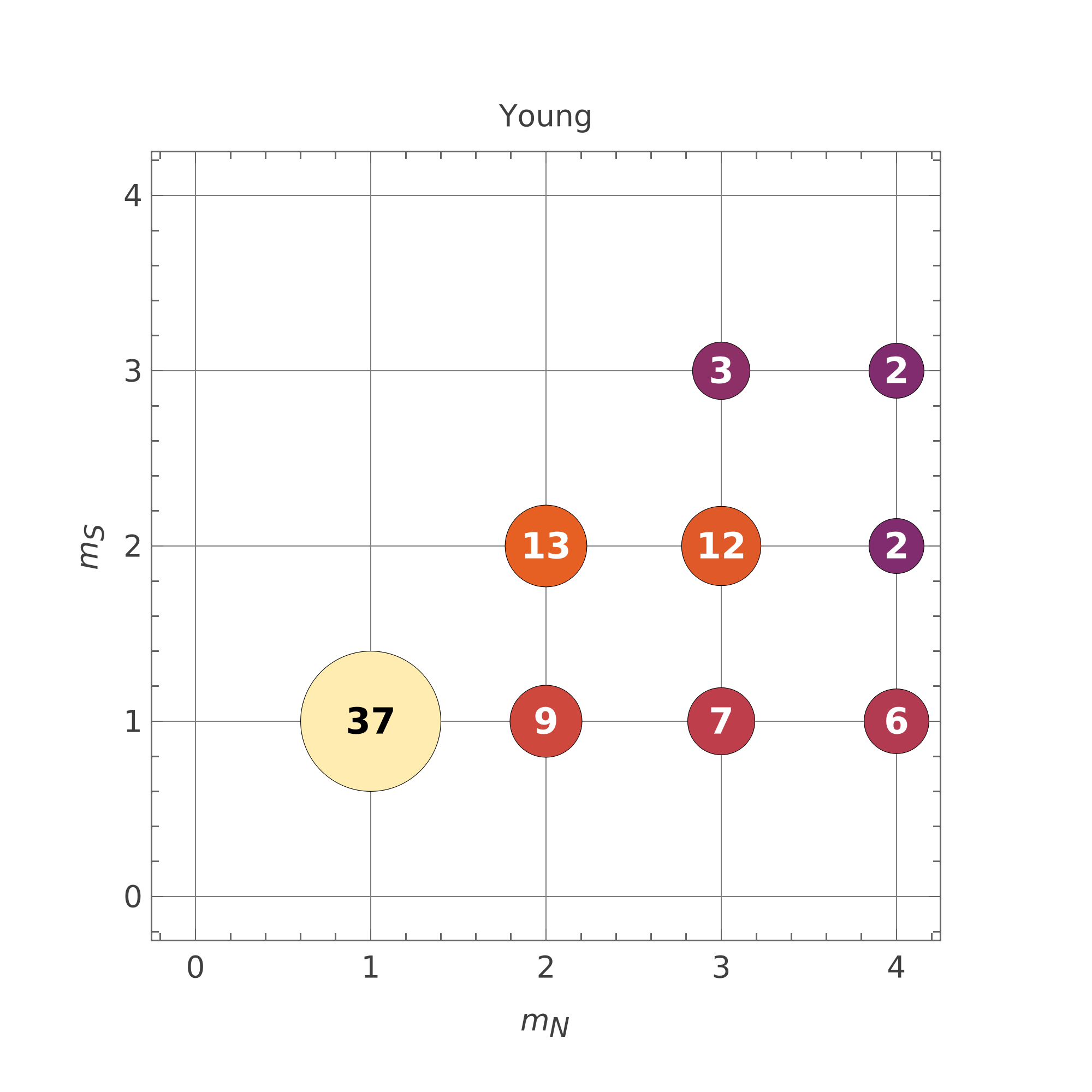} \\
	\includegraphics[width=0.9\linewidth]{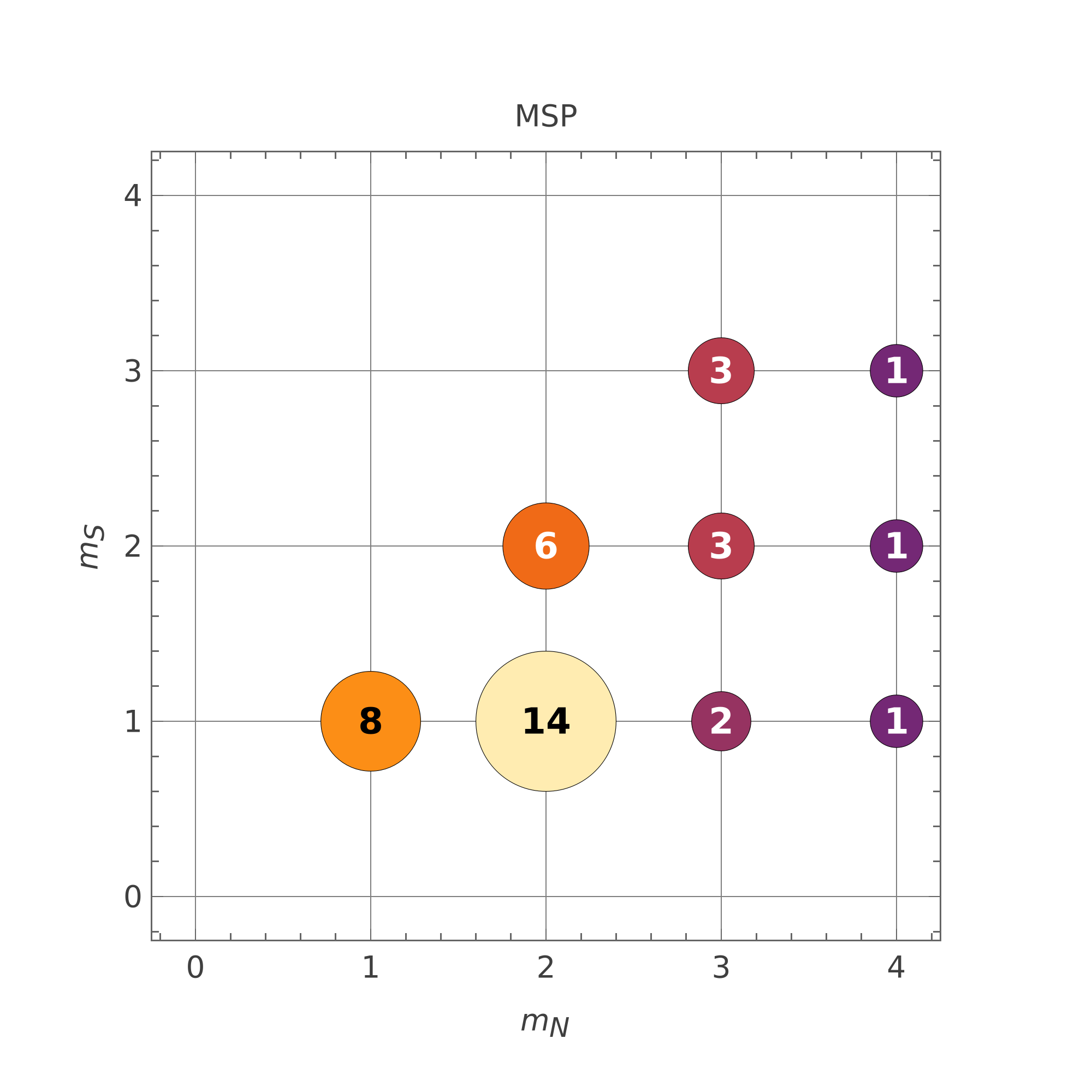}
	\caption{Distribution of the minimal modes $(m_{\rm N}, m_{\rm S})$ required to reasonably fit separately the young pulsars, top panel, and the millisecond pulsars, bottom panel.}
	\label{fig:synthese_mnx_msx_minmax_q5_jeunes_msp_mpos}
\end{figure}

In a real magnetosphere, the magnetic field is at least dipolar, corresponding to $m=1$ modes in the current density pattern for high inclination angle~$\alpha$. Solutions with at least one $m_{\rm x}=0$ should then be discarded based on the plasma flow pattern along the field lines (if we assume large obliquities $\alpha$ because for aligned rotators with $\alpha=0$, the pattern would be axisymmetric with $m=0$ and no pulsed emission is expected). Here we kept the $m=0$ as the split monopole allows for these modes. Fig.~\ref{fig:synthese_mnx_msx_minmax_q5_jeunes_msp_mpos} shows the correction in the counts brought by this additional constrain. The mode dominance remains largely unchanged confirming the dipolar quadrupolar segregation between young and millisecond pulsars.

\subsection{Comparison with other emission models}

Compared with other magnetospheric emission models such as the polar cap, slot gap, or outer gap scenarios, our current sheet model with asymmetric loading offers several advantages. It places the dominant emission region outside the light cylinder, consistent with the pulsar striped wind high energy emission model, and reproduces a wide range of observed light curve morphologies with only a few free parameters, making it both economical and flexible. The model provides a unified geometrical framework capable of describing both symmetric and asymmetric pulse profiles, while maintaining analytical simplicity and low computational cost, which facilitates efficient exploration of a huge parameter space. A key strength is the possible but not yet confirmed link between plasma loading parameters and light curve morphology, offering clear physical intuition and straightforward interpretability.

These advantages are nonetheless balanced by important limitations. The emissivity modulation remains phenomenological and is not derived from a self-consistent treatment of pair creation and injection. The model also lacks a detailed microscopic description of particle acceleration, distribution functions, and radiation processes, and should therefore be regarded as a geometrical, parametric framework rather than a completely physical model. In addition, its applicability is restricted by the assumed magnetic field structure, namely a split monopole configuration, which may affect the resulting light curve morphology and limits realism compared to global dissipative or fully kinetic simulations using a dipolar field.

\section{Conclusions\label{sec:conclusion}}

We have constructed an exhaustive atlas of $\gamma$-ray pulsar light curves based on split-monopole current sheet geometries, together with spatially dependent emissivity prescriptions that break both north-south and azimuthal symmetries. The model reproduces a broad range of observed $\gamma$-ray light-curve morphologies, including asymmetric and multi-peaked profiles, using only a few free parameters. Asymmetric plasma loading in the striped wind current sheet offers a natural explanation of the diversity of observed pulse shapes. The current sheet scenario thus appears to be a promising and physically motivated framework for high-energy pulsar emission.

Compared to symmetric current sheet models, the inclusion of asymmetric loading significantly improves the reproduction of several observed features. In particular, it allows one to account for asymmetric double-peaked profiles, multi-peaked light curves, and broad trailing components that are difficult to obtain in symmetric configurations. For instance, a north-south loading imbalance can shift the relative peak intensities, while azimuthal modulation can generate additional substructure or broaden one side of the pulse profile.
Within the adopted emissivity prescription, the vast majority of the observed profiles can be reproduced with low-order dipolar or quadrupolar like loading patterns, while more complex structures are required for only a small minority of objects. The diversity of $\gamma$-ray curves is not arbitrary, but results from a structured relationship between magnetospheric geometry, current distribution and the observer’s viewpoint.

Taken together, all these results suggest that the majority of $\gamma$-ray pulsar light curves are compatible with relatively simple plasma-loading structures, while millisecond pulsars (MSPs) exhibit a statistical preference for more complex by higher-order emissivity patterns. Within the adopted split-monopole geometry and emissivity prescription, the majority of observed $\gamma$-ray light curves are compatible with relatively low-order loading patterns, whereas MSPs show a statistical preference for higher-order patterns. This difference may be consistent with a larger contribution from non-dipolar magnetic structures in MSPs, but it does not yet constitute direct evidence for magnetic multipolarity, because the inferred loading complexity remains degenerate with the viewing geometry and with the assumed emissivity prescription. 

Future work needs to couple the present framework with global magnetospheric simulations in the force-free, dissipative, or at best in the particle-in-cell regimes in order to constrain the plasma-loading parameters from first principles. 
Finally, the inclusion of spectral and polarimetric observables should provide additional discriminants and help to further test the robustness of the model.



\appendix

\section{Light curve fitting results: full sample}
\label{app:full_sample}

\begin{figure*}[h]
	\centering
	\includegraphics[width=\linewidth]{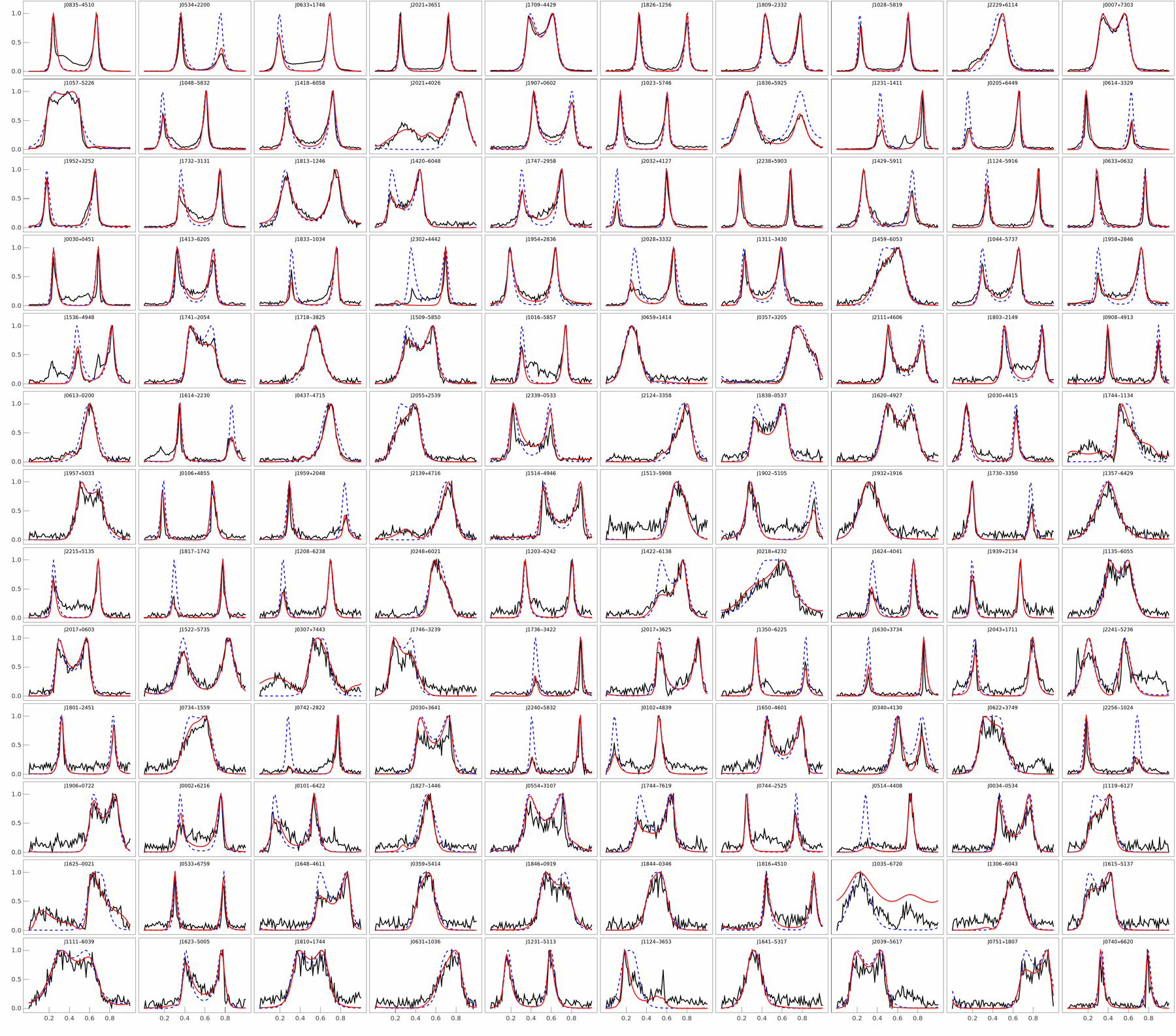}
	\caption{The full sample of light-curves fitted with the symmetric case $m_{\rm N}=m_{\rm S}=0$ in dashed blue lines and with the $m_{\rm N}=m_{\rm S}=1$ mode in solid red lines.}
	\label{fig:meilleur_ajustement_mn1_ms1_q5_asymx_echantillon}
\end{figure*}

\begin{figure*}[h]
	\centering
	\includegraphics[width=\linewidth]{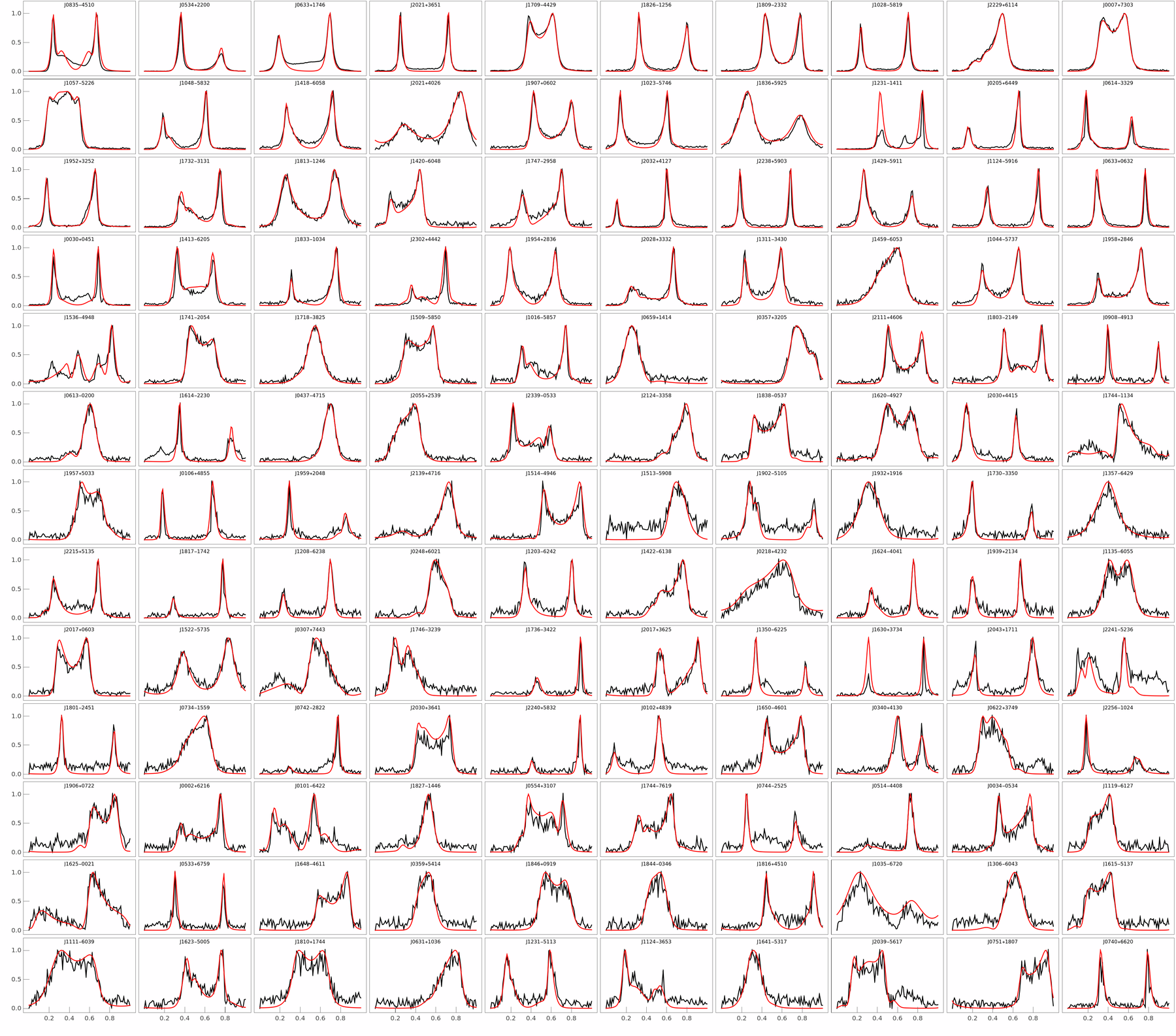}
	\caption{The full sample of light-curves fitted with the best mode satisfying $m_{\rm N} \leq 4$ and $m_{\rm S} \leq 4$ in solid red lines.}
	\label{fig:meilleur_ajustement_mnx_msx_q5_asymx_echantillon}
\end{figure*}

In this appendix, we show the full sample of fitted light curves, concerning the 130~pulsars we selected. Fig.~\ref{fig:meilleur_ajustement_mn1_ms1_q5_asymx_echantillon} shows the light-curves fitted with the symmetric case $m_{\rm N}=m_{\rm S}=0$ in dashed blue lines and with the $m_{\rm N}=m_{\rm S}=1$ mode in solid red lines whereas Fig.~\ref{fig:meilleur_ajustement_mnx_msx_q5_asymx_echantillon} shows the light-curves fitted with the best mode satisfying $m_{\rm N} \leq 4$ and $m_{\rm S} \leq 4$ in solid red lines.

\end{document}